\documentclass[12pt]{iopart}

\usepackage[utf8]{inputenc}
\usepackage[english]{babel}
\usepackage{eurosym}
\usepackage{xcolor}
\usepackage{mathrsfs}
\usepackage{graphicx}
\usepackage{float}
\usepackage{bbold}
\usepackage{bm}
\usepackage{cite}
\expandafter\let\csname equation*\endcsname\relax 
\expandafter\let\csname endequation*\endcsname\relax 
\usepackage{amssymb,amsfonts, amsmath}
\usepackage{textcomp}
\usepackage{hyperref}
\usepackage[T1]{fontenc}
\usepackage{lmodern}
\usepackage{csquotes}
\usepackage{siunitx}
\usepackage{caption}
\usepackage{subcaption}
\usepackage[normalem]{ulem}
\begin{document}

\title[]{A Framework for Synthetic Power System Dynamics}

\author{Anna Büttner$^1$, Anton Plietzsch$^{1,2}$, Mehrnaz Anvari$^{1,3}$ and Frank Hellmann$^1$}
\address{$^1$ Potsdam-Institute for Climate Impact Research}
\address{$^2$ Fraunhofer Research Institution for Energy Infrastructures and Geothermal Systems}
\address{$^3$ Fraunhofer Institute for Algorithms and Scientific Computing}

\ead{buettner@pik-potsdam.de}
\vspace{10pt}

	The paper is published in Chaos \cite{buettner_paper}. Please refer to the Chaos version from now on.\\
	
	Anna Büttner, Anton Plietzsch, Mehrnaz Anvari, Frank Hellmann; A framework for synthetic power system dynamics. Chaos 1 August 2023; 33 (8): 083120. \url{https://doi.org/10.1063/5.0155971}

\begin{abstract}
    We present a modular framework for generating synthetic power grids that considers the heterogeneity of real power grid dynamics but remains simple and tractable. This enables the generation of large sets of synthetic grids for a wide range of applications. For the first time, our synthetic model also includes the major drivers of fluctuations on short-time scales and a set of validators that ensure the resulting system dynamics are plausible. The synthetic grids generated are robust and show good synchronization under all evaluated scenarios, as should be expected for realistic power grids. A software package that includes an efficient Julia implementation of the framework is released as a companion to the paper.
\end{abstract}

    Power systems will be reconfigured as conventional generation is replaced by renewable energy sources (RES). The latter are often connected to the grid via inverters. The exact dynamical behavior and especially the stability of these inverters-based networks is not well understood. Thus, the availability of adequate synthetic power system models remains limited. However, it is vital to simulate future power grids to verify that this transformation does not result in undesired effects and blackouts. We introduce a framework for realistic synthetic power systems that can be used to study collective dynamical effects. We combine established methods such as realistic grid topologies and active power set-points. This framework opens new avenues for predicting the stability of future power grids using advanced techniques such as graph neural networks. 

\section{Introduction} 
    
    Synthetic power grids have become an important tool for studying the dynamics of power systems. Traditionally, most dynamical simulation studies in the engineering literature were performed using benchmark test cases, such as the "New England" IEEE 39-Bus System~\cite{athay1979practical} or the IEEE Reliability Test System-1996~\cite{grigg1999ieee}. The advantage of this approach is that models and parameters can be specified in great detail and the test cases are therefore highly realistic. Further, the use of standardized benchmark test cases guarantees comparability of different dynamic models and analytical methods. However, for many emerging research questions this approach can be quite limiting and the use of automatically generated synthetic grid models might be beneficial. This is for instance the case when the power system in a specific region should be studied but the detailed topology and parameters of the real grid are not publicly accessible. Often there is enough data or knowledge available to generate a synthetic grid that resembles the main properties of a real grid to a reasonable degree. An example is the algorithm by Birchfield et al.~\cite{birchfield_topology_2017,birchfield_structure_2017} that generates realistic transmission network topologies from spatial load distributions based on geographic population data. The algorithm is expanded in \cite{xu_dynamics_2017} to also enable transient stability analysis of the synthetic power grids. Besides the transmission system, synthetic grids are also required for studying mid- and low-voltage grids as their exact structure is often unknown~\cite{hoflich2012dena}. For German medium and low-voltage grids, the \emph{DingO} model \cite{dingo_Amme_2018} is an extensive and well-documented option to generate topologies \cite{dingo_Amme_2018} and supply and demand distributions \cite{electricity_consumption_Hülk_2017}. \emph{DingO} is part of the larger research project \emph{open eGo} and is open-source software that uses freely available data.

    Another important use case for synthetic power grid models is to generate large data sets of synthetic test cases that can be used to investigate the system dynamics with methods of machine learning~\cite{che2021active,nauck2022predicting}. A number of studies have shown that the network topology of grids has a direct influence on their dynamic stability~\cite{menck2014dead, schultz2014detours, kim2015community, kim2016building, hellmann2016survivability, nitzbon2017deciphering, kim2019structural}. However, most of these studies are based on simplistic component models and unrealistically homogeneous parameters, even though it is known that heterogeneities play an important role \cite{wolff2018power}. Graph-Neural-Networks have been shown to be a powerful method that could potentially extend these stability analyses to more realistic power grid models~\cite{nauck2022predicting,nauck_towards_2023, naucktowards2022}. The training of such neural networks requires large data sets of realistic grids, that are for example generated by a synthetic grid model.
    
    Finally, synthetic grid models will be crucially important for the investigation of the dynamic effects of future power grids. Within the next decades, the power system will undergo a fundamental transformation as new transmission infrastructure is built and conventional machines are replaced by renewable energy sources (RES). A major challenge is that the exact dynamical behavior of generation units is widely unknown as renewable generation units are connected to the grid via inverters with various control schemes. In order to maintain stability in such inverter-based grids, a certain share of these controls must be grid-forming. Today, most RES are still equipped with grid-following control schemes and hence, there is a lack of practical knowledge on the collective dynamical behavior of a large number of grid-forming generation units. It is therefore of great importance to do simulation studies of these systems to ensure that new technology being integrated into the grid does not lead to unexpected collective effects and blackouts~\cite{christensen2020high}. Unfortunately, there is a lack of both benchmark test cases as well as synthetic power grid models for studying such inverter-based grids.
    
    In this paper, we present a modular framework for generating synthetic grids that are suited for dynamic power system studies. We give an overview of all necessary steps from the generation of grid topologies, to the definition and parametrization of component models and the calculation of the steady state. The paper is accompanied by a software repository that provides an implementation of all algorithms described in this paper. Our approach is modular in the sense that users can easily adapt each step in the grid generation process to their own needs, e.g. by providing their own specific grid topologies or by using different dynamic models for the generating units in the system.  We focus on extra high voltage (EHV) level transmission grids, which in the continental European transmission grid includes the \SIrange{380}{400}{kV} and the \SI{220}{kV} voltage levels. Collective dynamical effects are traditionally studied in the highest grid layer \cite{review_dynamics_power_grids}, which is why we can rely on a comprehensive foundation there. In principle, the approach presented here can be extended to all grid layers. 
    
    The framework is designed to be capable of efficiently generating large numbers of synthetic grids with very limited input data. At the same time, the component models and parameters have a comparatively high level of realism: Generator and inverter models feature voltage dynamics, the active power production, and demand are heterogeneous and the parametrization of line admittances is according to data of the German transmission grid. The framework is therefore well-suited for applying machine learning methods, e.g. to predict dynamical stability from the structural properties of the grid.
    
    Another important feature is the possibility of modeling power grids with high shares of inverter-based generation units. For this, we bypass the problem that the exact dynamical models of such systems are still uncertain by using a technology and control scheme neutral model \cite{kogler_normal_2022} that has been shown to reproduce the behavior of a large class of different inverter controls as well as synchronous machine models. However, we also point out open research questions for improving the modeling of future power grids.

\section{Synthetic Power Grid Framework} 
\begin{figure}[H]
        \centering
        \includegraphics[width=\textwidth]{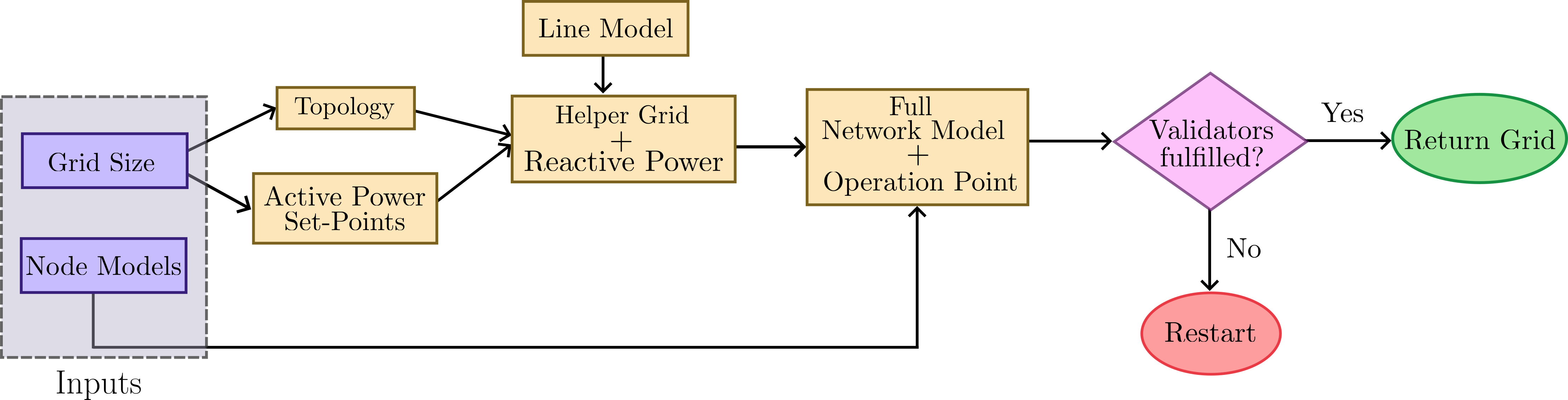}
        \caption{The default structure of the software framework. The user only needs to input the dynamic model of the nodes and the size of the power grid. Further steps, such as the generation of the topology and the power flow in the network are performed automatically. Before the power grid is returned by the software its behavior is validated to fulfill the stability criteria of real power grids. This flow chart only shows the currently implemented default structure, however as the framework is modular further options can be added over time.}
        \label{fig:flow_chart}
    \end{figure}

    For this project, we have chosen a framework to structure the synthetic power grid generation process. A framework in software development is defined as a semi-complete code basis that provides a reusable structure to share among applications \cite{tahchiev_junit_2011}. Users can integrate the framework into their own software and extend it to include specifically needed functionalities. The modularity and expandability of frameworks are needed for this project as researchers are interested in various properties and effects common in power grids which can be included in the framework over time. Furthermore, as more information on the structure of power grids under renewables becomes available it can easily be included in the existing software. As typical for frameworks we have developed a default structure that can be employed immediately by users. As the framework is modular each step can be interchanged as long as it adheres to the general structure. The default structure of the framework is shown in figure \ref{fig:flow_chart}.  In the default structure, the first step is to generate a topology or network structure for the synthetic power grid. Then active power set points for the nodes in the network are defined. The next step is to specify the node and line models in order to populate the networks with dynamics. Then an operation point, that fulfills certain stability criteria, is determined. In the last step, we validate the synthetic grids and ensure that the dynamic network properties are similar to those of real power grids that are carefully planned. 

    Most of the steps presented here have been used and validated individually in research projects before, however, they are now, for the first time, combined as a comprehensive package that is available for further research. Particularly, it is the first step towards a synthetic model of future power grids with high integration of RES. Each section contains a summary of a step in the framework as well as a critical analysis of the state-of-the-art. In the respective sections, we give an outlook and show which additional work could be done to improve the model, particularly for the representation of future power grids.
    
    For the analysis of the resulting grids, we also provide stochastic models that characterize fluctuations processes that are typical at the timescale of interest.
    
    \subsection{Grid Topology}
    The default topologies in our framework are generated using the random growth algorithm introduced in \cite{schultz_random_2014}. We choose this model as it is conceptually straightforward to generate a large number of interesting and plausible topologies and as it has little computational complexity, which is convenient for generating large ensembles of synthetic test cases. However, it is at the conceptual end of the synthetic grid spectrum. If the interest is to study dynamics on more realistic topologies, other models should be employed.
    
    The random growth algorithm \cite{schultz_random_2014} generates synthetic networks that resemble real-world EHV power grids with respect to the exponentially decaying degree distribution and the mean degree. The algorithm includes first an initialization phase, where a spatially embedded minimum spanning tree is generated, and then a growth phase. The growth phase makes use of a heuristic target function for the trade-off between the total line length, which determines the costs, and the smallest number of edges that would need to be removed to disconnect the grid into two parts, which influences the redundancy.
    
    The default parameters of the growth algorithm have been set to $[N_0, p, q, r, s] = [1, 1/5,  3/10, 1/3, 1/10]$, as employed in \cite{nitzbon2017deciphering}, where $N_0$ is the initial number of nodes in the minimum spanning tree, $p$, $q$ are the probabilities for generating a new redundant line, $s$ is the probability of splitting an existing line, and $r$ is the exponent for the trade-off between redundancy and cost. 
    
    Since distribution grids typically exhibit rather different network structures (mostly radial and ring topologies \cite{dingo_Amme_2018}) these parameters have to be adapted when the growth algorithm should be used for modeling lower voltage levels.
    
    For the default step, we assume that there is no correlation between the grid topology and the positioning of generation units in future grids. We thus assume that the transmission system topology will remain very similar to today, even if the position of generation units will be correlated to the renewable energy potentials and the location of the generation thus changes. This may not be entirely realistic and future studies should consider that the grid will be expanded and adapted to the new supply sources. However, such changes are expensive and time-consuming \cite{TYNDP_2020} and thus likely to be limited. To properly incorporate these aspects a synthetic geographical model, potentially incorporating economic optimization, such as \cite{brown2018synergies} is needed.

    \subsection{Active Power Distribution}
    In order to correctly represent the dynamics of the power grid, a realistic distribution of power in the grid is required. For this purpose the \emph{ELMOD-DE} \cite{Egerer2016Open} data set, an open-source spatially distributed, nodal dispatch model for the German transmission system is consulted. This data set has been chosen as it contains real data on demand and generation and has been accumulated from reliable sources such as the German Transmission Operators and the European Network of Transmission System Operators for Electricity (ENTSO-E). The \emph{ELMOD} data set represents the current load and capacity distribution, which means that RES are still in the minority. The analysis shown here is suitable for the distribution of active power in synthetic grids which should represent the status quo as most buses are either generation-heavy or load-heavy. Following \cite{taher2019enhancing}, which also analyses the data set, we examine the net power $\Delta P$ at each node given in the data set. 
    
\begin{figure}[H]
        \centering
        \includegraphics[width=0.7\textwidth]{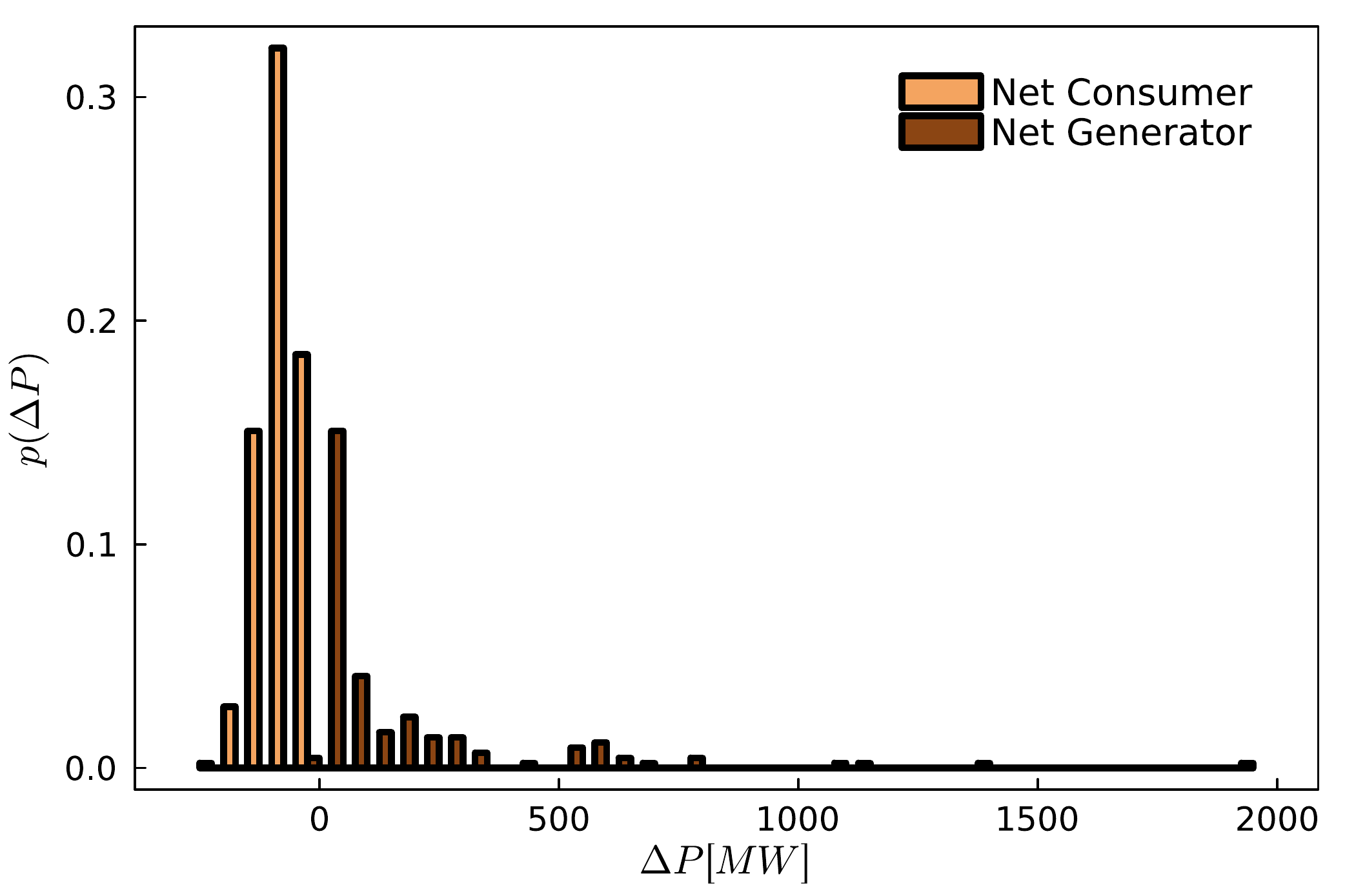}
        \caption{Histograms of the net nodal generation and consumption in the \emph{ELMOD-DE} \cite{Egerer2016Open} data set during the off-peak scenario. The distribution is bimodal and asymmetric. The power generation shows a heavy tail with a small share of net producers that generate more 750 \SI{750}{MW}.}
        \label{fig:power_distibution}
    \end{figure}

    The \emph{ELMOD} data set includes a time series for the total demand $P_{tot}$ in all of Germany. The demand is distributed to the individual nodes by introducing the nodal load share $ls_m$ which specifies the proportion of the consumption of a node $m$ from the total demand $P_{tot}$. It is distinguished between two different types of load scenarios, off-peak and on-peak. Egerer et al. \cite{Egerer2016Open} define on-peak and off-peak as the highest and lowest load level meaning the maximum and minimum of $P_{tot}$ respectively. The data set gives the load shares $ls_m$ for both scenarios the off-peak and on-peak. In the following, we will always work with the off-peak scenario. The consumption at a node $P_{con,m}$ is then given by:
    \begin{align}
        P_{con, m} = P_{tot} \cdot ls_m.
    \end{align}
    The \emph{ELMOD} data set includes the installed capacity for each generation unit $c_m$, which is the maximum power output the unit $m$ can produce. As multiple power plants can be connected to a single node, the nodal capacity $C_m$ is given by the sum of all capacities at the node $C_m = \sum_m c_m$. Typically, the full capacity of a generation unit is not available. In addition to the approach by Taher et al.~\cite{taher2019enhancing}, we also include the availability factors $a^{tech}$ for each technology during the off-peak scenario. The nodal availability $A_m$ is then given by: 
    \begin{align}
        A_m = \sum_m c_m \cdot a^{tech}.    
    \end{align}
    The total available power is defined as $A_{tot} = \sum_m A_m$.
    As there is no data about how much power each node generates at a given time point we follow the approach given in \cite{taher2019enhancing} and reduce the nodal availability $A_m$ by the factor $x = \frac{P_{tot}}{A_{tot}}$, such that generation and consumption are balanced. The nodal generation $P_{gen,m}$ is thus given by: $P_{gen,m} = A_m \cdot x$. Finally, we can define the net nodal power $\Delta P_m$ as:
    \begin{align}
        \Delta P_m = P_{gen, m} - P_{con, m}.
    \end{align}
    Figure \ref{fig:power_distibution} shows the distribution of the net nodal powers $\Delta P$ as a histogram. It can be seen that the distribution is bimodal and asymmetric and that the power generation is heavy-tailed. The heavy tail in the power distribution can be explained by the structure of today's power grid where the power is mostly produced by a small number of large generators. In the \emph{ELMOD} data set 301 nodes are classified as net consumers, while only 137 are net generators. For a future RES-heavy scenario the capacities and availabilities should be replaced with a model for the deployment of wind and solar renewable resources.
    
    Following \cite{taher2019enhancing} the active power $P$ of each node is sampled from a bimodal distribution, given by:
    \begin{align}
        p(P) = \frac{1}{2 \sigma \sqrt{2 \pi}} \left( \exp{\frac{(P-P_0)^2}{2\sigma^2}} + \exp{\frac{(P+P_0)^2}{2\sigma^2}} \right) \label{eq:power_dist}
    \end{align}
    in this work we will use $P_0 = \overline{\Delta P_{380}} \approx$ \SI{131}{MW}.
    
    The topologies used here, mimic the extra high voltage \SI{380}{kV} transmission grids. All following calculations are performed in a Per-Unit-System (p.u.), meaning that an appropriate base power $P_{base}$ and base voltage $V_{base}$ have to be chosen. As this work only examines the highest voltage layer of the grid the base voltage is simply chosen as $V_{base} = \SI{380}{kV}$. To define the base power for the \SI{380}{kV} level we extract all nodes that are connected to \SI{380}{kV} lines and calculate the mean $\overline{\Delta P_{380}} \approx $ \SI{131}{MW}. Based on the available data, we choose $P_{base} = \SI{100}{MW}$ as the base power for the synthetic power grids.
    
    For this work, we will adopt the bimodal model which was introduced in \cite{taher2019enhancing}. How this distribution will change due to the increasing share of RES but also changing consumption remains an open research question. A promising possibility is to base the distribution of active power supply on the renewable potentials of geographical areas. For this purpose, established software packages, such as atlite \cite{Hofmann_2021_atlite}, could be consulted. For the consumption side, new sectors with additional loads will be connected to the electric grid, for example, electric cars or hydrogen production.
    
    Furthermore, it should also be taken into account that the set points for the power change in the grid over time due to the evolution of the demand over the day and year. Typically these set points are updated every 15 minutes based on a cost optimization procedure. It would be valuable to study a grid and its dynamics under different load scenarios. Moreover, the demand is not constant between two dispatch times, but fluctuates, for example, studied in \cite{anvari2022data}. In section,\ref{sec:demand_fluc} we will apply the model for realistic demand fluctuations, which has been derived in \cite{anvari2022data}, to our power grids. Future work could also consider that the generation is typically distributed via an optimal power flow calculation to find the optimal dispatch.
    
    \subsection{Power Grid Model}
    On the most abstract level, we will mathematically describe power grids as systems of differential-algebraic equations (DAEs). The algebraic constraints are most commonly introduced via the load models, but can also appear when several generation units are present at a bus. Explicit DAEs are defined as:
    \begin{align}
         \dot x &= f(x, y) \label{eq:ODE} \\
         0 &= g(x,y)  \label{eq:constraint}
    \end{align}
    where equation \eqref{eq:ODE} and \eqref{eq:constraint} represent the differential and algebraic equations respectively. The vector $x$ holds the differential variables, whose derivatives appear in the DAE, while the vector $y$ gives the algebraic variables, whose derivatives do not appear.
    
    The specific models for the nodes and lines as well as for the networks are introduced in the following sections.
    \subsubsection{Node Models}
    \label{sec:nodal_models}
    Our synthetic grids will consist of grid-forming components, for example, power plants and novel types of inverters that contribute to grid stability and components without grid-forming capabilities, such as loads or grid-following inverters, that have to rely on an already stable grid. For this work, we have decided to use elementary nodal models to depict components with and without grid-forming abilities that are able to cover a large range of dynamical actors.
    
    In this work, PQ-buses \cite{Power_System_Dynamics_machowski} are used to represent the components without grid-forming behavior. The PQ-bus locally fixes the active and reactive power of node $m$:
    \begin{align}
         0 = (P_{set,m} + \mathrm{i} Q_{set,m}) - v_m \cdot i^{*}_m. \label{eq:PQNode}
    \end{align}
    where $P_{set,m}$ and $Q_{set,m}$ are the active and reactive power set points of the node, and $v_m$ and $i_m$ are the complex voltage and current of node $m$ which completely describe the physics of a balanced 3-phase AC system\cite{kogler_normal_2022}. The model can depict either loads or sub-networks of consumers and RES that are connected to the grid via grid-following inverters. The PQ-bus \eqref{eq:PQNode} is a constraint equation as given in equation \eqref{eq:constraint} and forces us to use the DAE description of the power grids.

    To represent grid-forming components we use the normal-form, a technology-neutral model for grid-forming actors, that has been introduced in \cite{kogler_normal_2022}. It has been shown that various models of grid-forming components, such as droop-controlled inverters~\cite{schiffer_model_2014} and synchronous machine models~\cite{schmietendorf_voltage_dynamics_2014}, can be expressed by the normal form. The normal form has been validated by numerical simulations and lab measurements of a grid-forming inverter so far, and work to identify normal form parameters for a wide range of grid forming actors is ongoing. A normal form at node $m$ with a single internal variable, the frequency $\omega_m$, is given by:
    \begin{align}
        \nu_m &= v_m v_m^* \nonumber \\
       \dot{\omega}_m &= A^{\omega, m} + B^{\omega, m} \delta \omega_m + C^{\omega, m} \delta \nu_{m} + G^{\omega, m} \delta P_m + H^{\omega, m} \delta Q_m \label{eq:normalform} \\
        \frac{\dot{v}_m}{v_m} &= A^{v,m} + B^{v,m} \delta \omega_m + C^{v,m} \delta \nu_{m} + G^{v,m} \delta P_m + H^{v,m} \delta Q_k \nonumber
    \end{align}
    where $v_m$ is the complex voltage. $\delta P_m$ and $\delta Q_m$ represent the difference between the active and reactive power to the set points. $\delta \nu_m$ is the difference of the squared voltage magnitude $\nu_m$ to the squared voltage set-point. The other coefficients are the modeling parameters that capture all the differences between the various models the normal form can represent. The parameters $A^{\omega, m}$ and $A^{v,m}$ are zero when the system is, as in our case, defined in the co-rotating reference frame. In the normal form, all structural differences between models are absorbed in the parametrization.

    The free parameters for the normal form can be gathered by approximating other models, moreover, it is also possible to derive them from experimental data, which has also been performed in \cite{kogler_normal_2022} for a specific type of inverter in a lab. For the example provided in this work, we will use a normal form approximation of a droop-controlled inverter \cite{schiffer_model_2014} whose parameters can be derived analytically. 
    
    The exact ability of the normal form to cover all needed dynamics is a subject of current research. Future work will include measurements on different types of inverters and deriving the parameters of the normal form from the data. This is a crucial step to study the dynamics and stability of realistic future power grids, which will consist of a variety of interacting grid-forming inverters.
    
    In addition, we use a slack bus \cite{Power_System_Dynamics_machowski} for the load flow calculation. The slack bus locally fixes the voltage $v_m$ of node $m$:
    \begin{align}
        0 = v_{set,m} - v_m \label{eq:slack}.
    \end{align}
    where $v_{set,m}$ is the set point voltage. The voltage magnitude $|v_{set,m}|$ of the slack is typically set to 1 p.u. and its voltage angle is $\phi_m = 0^\circ$. The slack bus is not included in the resulting dynamic synthetic power grid. It is only an ancillary component that is used as the reference for all other buses in the system while solving the load flow problem, as described in section \ref{sec:load_flow}. The active and reactive power of the slack bus are free to change to compensate for the power imbalance in the network. Therefore it is assumed that the slack bus has a large amount of energy stored which can be released quickly. The slack bus is typically considered to be a large power plant or battery, a connection point to a higher grid layer, or another part of the power system which is not modeled explicitly. 
    
    \subsubsection{Line Model}
    \label{sec:line_models}
    For this work, the Pi-Model, see for example \cite{power_system_lecture}, is used. In the Pi-Model the impedance $Z_{km} = \frac{1}{Y_{km}}$ is placed in the center of the line. The capacitance between the line and the ground is also taken into account by introducing the shunt admittance $Y_{sh, km}$ which is placed, in parallel, at both ends of the line.
    The current on the lines connecting node $k$ and $m$ is then given by \cite{power_system_lecture}:
    \begin{align}
        i_{km} &= Y_{km}(v_k - v_m) + Y_{sh, km} v_k \\
        i_{mk} &= Y_{km}(v_m - v_k) + Y_{sh,km} v_m
    \end{align}
    where $Y_{km}$ is the admittance of a line connecting node $k$ and $m$ and $Y_{sh, km}$ is the shunt admittance. $v_k$ and $v_m$ are the complex nodal voltages. Combining the nodal and line models we obtain the full network model. The current injected at node $k$ is given by:
    \begin{align}
        i_k = \sum_m i_{km}
    \end{align}
    and the power flow in the network is defined as:
    \begin{align}
        S_k = v_k i_k^* = P_k + \mathrm{i} Q_k
    \end{align}
    where $S_k$ is the apparent power at node $k$ and $P_k$ and $Q_k$ are the real and reactive power injected at $k$ respectively \cite{Power_System_Dynamics_machowski}.
    
    The impedance and shunts are calculated according to the \emph{dena} model of standard \SI{380}{kV} overhead power lines \cite{dena_dena-verteilnetzstudie_2012} given in table \ref{tab:overhead_lines_parameters}. The reactance per unit length $X$ is specified for the nominal frequency of \SI{50}{Hz}, which is why we use a static line model here. The total admittances are calculated according to:
    \begin{align}
        k_{c} &= \frac{c}{c_{t}}\\
        k_{w} &= \frac{w}{w_{t}} \\
        Y_{km} &= \frac{k_{c} k_{w}}{(R + j X) \cdot l_{km}}  \label{eq:line_admittance} \\
        Y_{sh, km} &= \frac{-(j \omega C_{sh})  k_{c} k_{w} }{2} \cdot l_{km}
    \end{align}
    where $l_{km}$ is the line length in kilometers. For consistency, we fix the grid frequency $\omega$, in the shunt admittance $Y_{sh, km}$, to the nominal frequency. The coefficients $k_c$ and $k_w$ define the ratio between the typical number of cables $c_{t}$ and wires $w_{t}$ and the actual numbers of cables $c$ and wires $w$ in the line \cite{medjroubi_open_2017}. The typical numbers of cables and wires are 3 and 4 respectively for transmission lines in the \SI{380} level in Germany \cite{medjroubi_open_2017}. In the default version of the algorithm, we assume that all transmission lines have the typical number of cables and wires. In section \ref{sec:Probabilistic_Capacity_Expansion} we introduce an additional step in the algorithm where probabilistic power flow scenarios are considered. The line capacities are increased, by adding new cables to existing lines, if a load scenario leads to an overload.
     
    \begin{table}[H]
        \centering
        \begin{tabular}{|l|l|l|l|}
            \hline
            Voltage level & $R$ [\si{\Omega \per km}] & $X$ [\si{\Omega \per km}] & $C_{sh}$ [\si{nF \per km}]  \\ \hline
            \SI{380}{kV}         &         0.025            & 0.25                    & 13.7  \\ \hline
        \end{tabular}
        \caption{Standard overhead line parameters according to \cite{dena_dena-verteilnetzstudie_2012} for the typical number of cables and wires.}
        \label{tab:overhead_lines_parameters}
    \end{table}
    
    To calculate the line properties the lengths of the transmission lines are needed. As the model of \cite{schultz_random_2014} generates an embedded topology, but does not provide a spatial scale, we need an additional step to determine the spatial scale. This is done by requiring that the line lengths of the synthetic grids resemble the line lengths of real EHV grids.
    
    The line lengths $l_{mk}$ in kilometers are obtained by converting the Euclidean distances $d_{mk}$ of the lines, which are generated by the random growth model \cite{schultz_random_2014}. The conversion factor $c_l$ is given by the mean length $\langle l \rangle$ of overhead lines in the extra high voltage (EHV) level, that concerns voltages equal or greater than \SI{220}{kV}, divided by the mean euclidean distance $\langle d \rangle$: 
    \begin{align}
        c_l &= \frac{\langle l \rangle}{\langle d \rangle}   \\
        l_{mk} &= c_l \, d_{mk}. 
    \end{align}
    Additionally, we used the shortest line $l_{min}$ in the EHV level as a threshold. The admittances of lines that are shorter than $l_{min}$ are set to the threshold impedance of the shortest line.
    
    The mean line length was determined from the \emph{SciGRID} data set \cite{medjroubi_open_2017}, which consists of openly available geographic data of the German power grid. At the time of the creation of the data set the coverage of the EHV level in Germany was around 95\% \cite{medjroubi_open_2017}, which thus offers an excellent basis for such a study.
    
    The \emph{ELMOD} data-set \cite{Egerer2016Open} also offers a network topology that is based on network plans by the transmission system operators (TSOs) and OpenStreetMap data. Since the data in \emph{SciGRID} is better documented and the study deals much more intensively with the network topology, we base our transmission line lengths on \emph{SciGRID}. Still, for completeness, we will also analyze the data from \emph{ELMOD}. A comparison between \emph{SciGRID}, \emph{ELMOD}, and our synthetic grids, which are based on \emph{SciGRID} is given in table \ref{tab:transmission_line_lengths}.
    \begin{table}[H]
        \centering
        \begin{tabular}{|l|l|l|l|}
            \hline
                                                        & $\langle l \rangle$ [\si{km}] & $\sigma_l$ [\si{km}] & $l_{min}$ [\si{km}]  \\ \hline
            \emph{SciGRID}  \cite{medjroubi_open_2017}  & 37.13                    & 36.59                & 0.06                 \\ \hline
            \emph{ELMOD-DE}   \cite{Egerer2016Open}     & 40.98                    & 35.54                & 0.42                 \\ \hline
            Synthetic Grids                             & 37.13                    & 34.6                 & 0.06                 \\ \hline
        \end{tabular}
        \caption{Comparison of transmission line lengths between different models. The values for the synthetic grid were calculated by generating 10000 different topologies. The mean line length is given by $\langle l \rangle$, the standard deviation of the line length is $\sigma_l$ and the minimal line length by $l_{min}$.}
        \label{tab:transmission_line_lengths}
    \end{table}
    In table \ref{tab:transmission_line_lengths} it can be seen that the mean line length, as well as the standard deviation of the line length of \emph{SciGRID} and \emph{ELMOD}, match well. Furthermore, it can be seen that our synthetic grid line length shows a standard deviation that matches the \emph{SciGRID} as well as the \emph{ELMOD}. The most significant difference between the two data sets is the minimum line length $l_{min}$, which is about \SI{4}{km} in \emph{ELMOD} and about \SI{60}{m} in \emph{SciGRID}. For the reasons that were stated above, we have adopted $l_{min}$ from \emph{SciGRID}.
    
    Future work would also include not only analyzing the mean and standard deviation of the length but also matching the distributions of line lengths (see fig.~\ref{fig:line_lengths} in the Appendix). This goes beyond the random growth algorithm \cite{schultz_random_2014} which is currently used, and would require an algorithm that considers line lengths, node locations, and a spatial embedding. A preliminary study \cite{networks_of_networks} on extending the algorithm which uses different node positioning rules has been performed but it does not deal with recovering the correct line length distribution.
    
    \subsection{Operation Point and Reactive Power}
    \label{sec:load_flow}
    Finding a stable operation point for synthetic power grids is challenging as power systems are generally non-linear and multi-stable. The AC load flow has no guarantee for convergence. Even if it converges only the synchronous fixed points whose voltage magnitudes are all close to 1 p.u. are physically meaningful for power grids. Securing the voltages to be close to their nominal values is a difficult task to accomplish. Typically, the reactive powers of the nodes are adjusted to control the voltage magnitudes in the power grid \cite{Power_System_Dynamics_machowski}. However, for many synthetic grid models, there is no prior information about the reactive power flow.
    
    Reactive power planning is considered to be one of the most intricate problems in power grid planning \cite{zhang_reactive_2007}. The review article \cite{zhang_reactive_2007} gives an excellent overview of the objectives and constraints that are considered in reactive power planning. Instead of implementing one of the complex established models presented in \cite{zhang_reactive_2007} we use a straightforward method to solve the reactive power flow. We employ the voltage stability objective, which is also a standard objective according to \cite{zhang_reactive_2007}, and assume that it has to be met perfectly. This requirement uniquely determines the reactive powers at the nodes. 

    We generate an ancillary power grid with the same topology and line models as the full power grid. The ancillary power grid consists of PV buses where all nodes are constrained to have voltages magnitudes of $V_m = 1 p.u.$ and the same active power that they generate in the actual power grid. One of the nodes is randomly turned into the slack bus \eqref{eq:slack} of the system that accounts for any power imbalances, for example, due to line losses. The reactive powers of the ancillary grid are found by using the power flow calculation of \texttt{PowerModel.jl} \cite{power_models} and a root-finding algorithm to find a steady state. The operation point of the ancillary grid is used as the initial guess for the operation point search of the actual grid.

    As the synthetic grids generated in this work have less than 10000 nodes our approach still leads to feasible power flow solutions. Once the grids become bigger a more in-depth reactive power flow planning algorithm, such as \cite{reactive_power_planning}, will be needed to find feasible operation points. 
    
    \subsection{Validators}
    Real-life power grids are planned carefully to lead to stable operations. Synthetic processes can never fully capture this planning stage. Instead, we use a rejection sampling approach. Synthetic power grids whose dynamics do not satisfy the stability properties of real-life power grids are rejected. In this section, we introduce a set of validators that review the stability of the synthetic power grids in their operation point. To assess our default settings we generated a set of synthetic networks with different sizes and studied the number of rejections. We generated power grids ranging from 100 to 1300 nodes with a step size of 25 nodes. For each grid size, we generate 100 power grids and can report that no grid was rejected. 
    
    \subsubsection{Voltage Magnitude}
    Firstly, we verify that the nodal voltage magnitudes fulfill the standard of the EN 50160 report \cite{power_quality_standard}. The report specifies that the average 10 minutes root mean square voltage has to stay within the bounds of $\pm 10 \%$ for 95$\%$ of the week. We assure this by validating that all nodal voltage magnitudes are $V \approx 1 p.u.$ in the operation point. If the set points of the system and the parametrization have been chosen properly the voltage condition should be fulfilled. Even if the reactive power is chosen to ensure a stable power flow with good voltage magnitudes, incorrectly specified control dynamics or machine parameters, can still lead to a violation of the voltage conditions in the operating point. Thus even in this case, the verification of the voltage condition is still essential in order to catch such unrealistic parametrizations.
    
    \subsubsection{Line Loading Stability Margin}
    \label{sec:line_loading}
    In a stable operation of the power grid, no line is overloaded. There are different thresholds for the allowed loading of a transmission line. In this work, we focus on the threshold which is determined by the stability margin and depends on the physically possible limit of the line $P_{max}$.
    
    The power flow transferred over a line connecting node $m$ and $k$, neglecting the reactive power flow and line losses, is given by:
    \begin{align}
        P_{mk} = \frac{v_m v_k}{X_{mk}} \sin(\theta_{mk})
    \end{align}
    where $v_m$ and $v_k$ are the nodal voltage magnitudes, $X_{mk}$ is the line reactance and $\theta_{mk}$ is the difference of the voltage angles of node $m$ and $k$. The transferred power becomes maximal when $\theta_{mk} = \frac{\pi}{2}$. Thus the physically possible limit of the line is $P_{max} = \frac{V_m V_k}{X_{mk}}$. To assure a stable power system transmission lines are operated well below this limit and a so-called stability margin $sm$ is introduced \cite{transmission_systems_book}. The transferred power of a line $P_{rated}$ must therefore be below a threshold given by: $P_{rated} \leq P_{max}(1 - sm)$. In this study, we choose $sm = 0.3$ as suggested in \cite{transmission_systems_book}. If any line loading in our power grid violates this threshold we reject the power grid.

    \subsubsection{Small Signal Stability Analysis}
    Since the grids we consider in this work are described by DAEs, we cannot simply study the eigenvalues of the Jacobian in the equilibrium to determine the linear stability of the system. Instead, we perform a small signal stability analysis for DAEs according to \cite{milano_scripting}. 
   
    In this approach, the eigenvalues of the so-called reduced Jacobian, or state matrix $J_{red}$ are examined. The reduced Jacobian is set up by decomposing the full Jacobian matrix $J$ into the following blocks:
    \begin{align}
        J = 
            \begin{bmatrix}
            \partial_x f & \partial_y f\\
            \partial_x g & \partial_y g
        \end{bmatrix}
    \end{align}
    where $\partial_x f$ is an abbreviation for the matrix of partial derivatives of the right-hand side of the differential equations $f$ with respect to the differential variables $x$, and $\partial_y g$ gives the matrix of the partial derivatives of the algebraic equations $g$ with respect to the algebraic variables $y$.
    
    Following \cite{milano_scripting}, the reduced Jacobian is defined as:
    \begin{align}
        J_{red} &= \partial_x f - D \\
        D &= \partial_y f \left( \partial_y g \right)^{-1} \partial_x g
    \end{align}
    where $D$ is the degradation matrix. The eigenvalues of $J_{red}$ can be examined as usual again, meaning that power grids whose eigenvalues of $J_{red}$ have positive real parts are classified as linearly unstable. Power grids whose operation point is linearly unstable would not exist in reality and therefore have to be rejected before any further investigations are performed.
    
    \subsubsection{Probabilistic Capacity Expansion} 
    \label{sec:Probabilistic_Capacity_Expansion}
    So far we have only assured that the synthetic power grids are stable under a single power set point that was drawn from the probability distribution \eqref{eq:power_dist}, or any other source. However, in real power grids, the set points are updated regularly, e.g. in Germany, a new demand plan is implanted every 15 minutes. Therefore, it is important to also verify the stability of the grid under different set points. In principle, all validators can be applied to an ensemble of set points. In this work, we only focus on the capacity of lines, as this is the most directly affected by the demand, and assure that there is always enough line capacity to cover the expected load cases.
    
    We sample completely new set-points from the bimodal distribution \eqref{eq:power_dist} but double the mean power $P_0$ in order to study the system under more stress. A more realistic analysis of high-stress power flow scenarios would require an extensive investigation of the expected set points and is therefore beyond the scope of this paper.
    
    For each new scenario, we calculate the load flow in the grid and then analyze the line loading as given in section \ref{sec:line_loading}. If a line is overloaded we add three additional cables to the line to increase its admittance as in equation \eqref{eq:line_admittance}. This approach is repeated for $N$ different scenarios. So far no new cables were added for all performed simulations. This is to be expected since, in the \emph{SciGRID} \cite{medjroubi_open_2017} data set, more than 90$\%$ of the EHV transmission lines have the typical number of cables. It is nevertheless important to validate the grid under different load scenarios to assure its stability. Furthermore, this capacity evaluation could become important once more realistic load scenarios are evaluated, which in the future could include the weather-dependent time series generated by atlite \cite{Hofmann_2021_atlite}.
    
    While these validators cover the most basic functioning of the grid, further conditions can also be considered. A natural extension for future work would be to add N-1 stability as a condition that the grids need to satisfy.
\section{Nodal fluctuations} 
    \label{sec:node_fluctuations}
    Due to the increasing share of variable RES, i.e. wind and solar energy, power grids are exposed to new sources of fluctuations. RES are fluctuating at different time scales \cite{apt2007spectrum,woyte2007fluctuations} and, particularly, have intermittent fluctuations at short time scales \cite{anvari2016short}. Along with supply-side fluctuations, recent studies of high-resolution recorded electricity consumption demonstrate intermittent fluctuations on the demand-side \cite{anvari2022data,wright2007nature,monacchi2014greend} as well. To generate synthetic power grids that imitate the dynamics of real power systems at such short time scales, fluctuations have to be considered both on the supply and demand side.  
             
    Here we introduce the stochastic processes that generate fluctuating wind and solar power, as well as demand time series. These models have been derived to ensure that, these synthetic time series have the same short time-scale stochastic characteristics as empirically observed in real data. Therefore, one can confidently use the synthetic time series for further research in power grids, and consider the response of power systems to these fluctuations. The effects on the grid frequency are illustrated in section \ref{results}. 

\subsection{Supply fluctuations}
\label{sec:supply_fluc}
    The intermittent nature of wind speed and solar irradiance, along with their turbulent-like behavior, which transfers to wind and solar power and, consequently, to power grids has been widely discussed \cite{anvari2016short,apt2007spectrum,milan2013turbulent,curtright2008character}.
    As demonstrated in these studies, wind and solar power are non-Gaussian time series and have heavy-tailed probability distribution functions (PDF). Extreme fluctuations, such as a $90\%$ reduction in power in just a few seconds, occur often in RES. These fluctuations can present additional challenges for maintaining the stability of power systems. 

    Here, we employ a non-Markovian Langevin-type stochastic process \cite{schmietendorf2017impact}, as well as a jump-diffusion model \cite{anvari2017suppressing} to generate respectively wind and solar power with similar short time-scale characteristics as the empirical data sets.
    The Langevin-type model used here is:
    \begin{equation}
        \dot{P}_{wind}(t)={P}_{wind}(t)(\Gamma-\frac{P_{wind}(t)}{P_0})+\sqrt{\kappa P_{wind}^2(t)}n(t)
        \label{eq:LT}
    \end{equation}
    where, $\Gamma$ and $P_0$ are constant parameters, and $\kappa$ is a parameter with which one can tune the intensity of the noise $n$. The exact values of the parameters used in our simulations are given in section \ref{results}. The noise $n$ is obtained from the following Langevin equation:
    \begin{equation}
        \dot{n}(t)=-\gamma n(t)+\zeta(t)
        \label{eq:LN}
    \end{equation}
    where $\zeta$ is a Gaussian noise with $\langle \zeta(t)\rangle=0$ and $\langle\zeta(t)\zeta(t^{'})\rangle=\delta(t-t^{'})$. The jump-diffusion model emulating short time-scale fluctuations in solar power is:
    \begin{equation}
        dP_{solar}(t)=D^{(1)}(P_{solar},t)dt+\sqrt{D^{(2)}(P_{solar},t)}dw(t)+\eta dJ(t)
        \label{eq:JD}
    \end{equation}
    where $D^{(1)}$ and $D^{(2)}$ are respectively the drift and diffusion coefficients. In eq.~\ref{eq:JD}, $dw$ is the Wiener process and $dJ$ is the Poisson process with jump size $\eta$, which is assumed to be a normally distributed random number, i.e $\eta\sim N(0,\sigma_{\eta})$. The Poisson process comprises also a jump rate, which we call $\lambda$. The advantage of the jump-diffusion model is that it is a non-parametric model, i.e. all parameters are derived from the empirical data sets. 

\subsection{Demand fluctuations}
\label{sec:demand_fluc}
    Standard load profiles used to balance energy in the grid in advance have a time resolution of 15 minutes. Shorter time scales are balanced by control mechanisms rather than by trading. To study the dynamics at short time scales the load profiles are thus of limited use. Instead, we consider empirical measurements of loads that have a high enough resolution to reveal short-term fluctuations such as \cite{anvari2022data,wright2007nature,MarszalPomianowskaetal2016}.
    
    Here, we apply the superstatistics model introduced in \cite{anvari2022data} to generate the short time-scale fluctuations of the demand side. Following the superstatistical approach, the demand fluctuations are obtained by taking the 2-norm of several Gaussian distributions plus a constant offset $\mu_{MB}$:
    \begin{equation}
        P^{fluc}(t)=\sqrt{(z_1(t))^2+(z_2(t)^2)+...+(z_J(t))^2}+\mu_{MB}
    \end{equation}
    where we use $J=3$ as discussed in \cite{anvari2022data}, and $z_i(t)$ is obtained from the following Langevin equation:
    \begin{equation}
        dz_i(t)=\gamma z_i(t)dt+\epsilon dw_i    
    \end{equation}
    where $dw_i$ is the Wiener process with a mean $0$ and standard deviation $\sigma=\epsilon/\sqrt{2\gamma}$. We employ the same parameter values $\mu_{MB}$, $\gamma$, and $\epsilon$ as reported in \cite{anvari2022data}.
    
    It should be noted that the stochastic time series we have introduced here is based on empirical measurements of power grid actors that are typically not directly connected to the highest level of the power grid. As not all producers and consumers connected to a bus are perfectly correlated the fluctuations would be attenuated in reality. Unfortunately, few or no measurements of the actual correlations of fluctuations exist, which is why we need to leave this point to future work.

\section{Simulation Examples} 
\label{results}
\begin{figure}[H]
         \centering
         \includegraphics[width=0.7\textwidth]{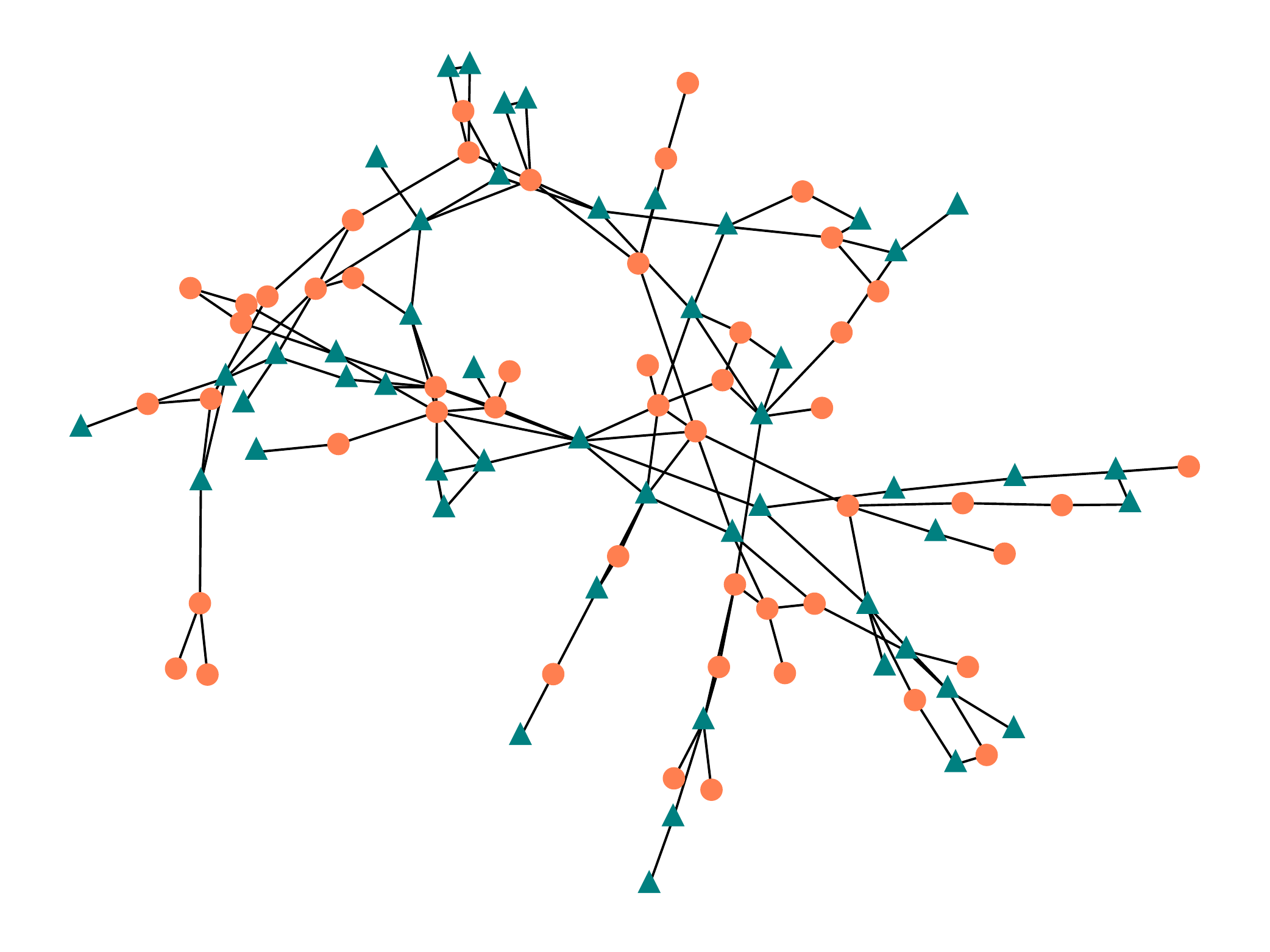}
         \caption{Network structure of a synthetic power grid. Triangular and circular nodes depict grid-following and grid-forming inverters respectively.}
         \label{fig:network}
    \end{figure}

    In this section, we generate a fully electrified synthetic power grid, whose structure is shown in figure \ref{fig:network}, and study its behavior in response to the three different fluctuations processes that have been introduced in section \ref{sec:node_fluctuations}. The synthetic grid that we consider here consists of 100 nodes with an equal share of grid-following and grid-forming inverters. We expect that future power grids will have a high share of variable renewable energies and, therefore, we consider multi-node fluctuations in this example. We assume that the grid-forming inverters are equipped with sufficiently large storage units. Hence the RES fluctuations are only fed into the grid via the grid-following inverters.
    
    The fluctuations $P_{fluc,i}(t)$ are added to the set points $P_{set,i}$ of the nodes. This results in the following equation for the active power $P_i$ at node $i$:
    \begin{align}
        P_i(t) = P_{set,i} + P_{fluc,i}(t).
    \end{align}
    For the different processes, we will analyze the two edge cases, completely correlated fluctuations, meaning that all nodes have the same fluctuating time series $P_{fluc}(t)$, and secondly, completely uncorrelated fluctuations where all nodes have different fluctuating time series. 
    
    In order to compare the results we will study two performance measures, the synchronization norm $||\mathcal{L}||_{sync}$ \cite{andreasson2017coherence} and the $L_2$ norm of the average deviation from the nominal grid frequency $|| \mathcal{L} ||_{dev}$ \cite{plietzsch_linear_response_2019}: 
    
    \begin{align}
        ||\mathcal{L}||_{sync} &= \sqrt{ \frac{1}{T}\int_0^{T} \frac{1}{N} \sum_{m=1}^{N} \left( \omega_m(t) -  \frac{1}{N} \sum_{k=1}^{N} \omega_k(t) \right)^2 dt} \label{eq:sync_norm} \\
        ||\mathcal{L}||_{dev} &= \sqrt{ \frac{1}{T}\int_0^{T} \frac{1}{N} \sum_{m=1}^{N} \left( \omega_m(t) - \omega_0 \right)^2 dt}
    \end{align}
    where $\omega_0$ is the nominal grid frequency. The indices $m,k$ run over all $N$ grid-forming inverters as the grid-following inverters have no internal frequency dynamics \eqref{eq:PQNode}. 
    
    The synchronization norm \eqref{eq:sync_norm} measures the synchronicity in the power grid. A large synchronization norm expresses a lack of synchronization. The synchronization norm however neglects any fluctuation of the so-called bulk \cite{zhang2019fluctuation}, the joint response of the entire power grid, of synchronous frequencies. Therefore the authors of \cite{plietzsch_linear_response_2019} introduce the deviation norm $|| \mathcal{L} ||_{dev}$ which measures the contribution of the bulk to the fluctuations. In \cite{plietzsch_linear_response_2019} it has been shown that the bulk is the dominant contributor in response to single-node renewable energy fluctuations. 
    
    The results are summarized in table \ref{tab:performance_measures_corr} and \ref{tab:performance_measures_uncorr}. In all cases, it can be seen that the deviation norm $|| \mathcal{L} ||_{dev}$ is larger than the synchronization norm. This indicates that the bulk fluctuations are the main contributors to multi-node renewable energy fluctuations as well. This holds for all fluctuation processes and for both edge cases, the correlated or uncorrelated fluctuations. Furthermore, we can see that the deviation norm is smaller for the uncorrelated case than for the correlated case, which is to be expected. Moreover, it can be seen that the synchronization norm is very small for all cases which implies that the networks have a high degree of synchronicity under renewable energy fluctuations.  
    
    \begin{table}[H]
    \centering
        \begin{tabular}{|l|l|l|}
            \hline
                                & $||\mathcal{L}||_{sync}$ & $||\mathcal{L}||_{dev}$ \cite{plietzsch_linear_response_2019} \\ \hline
            Wind Fluctuations   & 0.001     & 0.874     \\ \hline
            Demand Fluctuations & 0.002     & 1.952     \\ \hline
            Solar Fluctuations  & 0.033     & 0.686     \\ \hline
        \end{tabular}
        \caption{Performance measures for completely correlated fluctuations.}
        \label{tab:performance_measures_corr}
    \end{table}
    
    \begin{table}[H]
    \centering
        \begin{tabular}{|l|l|l|}
            \hline
                                & $||\mathcal{L}||_{sync}$ & $||\mathcal{L}||_{dev}$ \cite{plietzsch_linear_response_2019} \\ \hline
            Wind Fluctuations   &  0.001    & 0.153     \\ \hline
            Demand Fluctuations &  0.002    & 0.417     \\ \hline
            Solar Fluctuations  &  0.027    & 0.099     \\ \hline
        \end{tabular}
        \caption{Performance measures for completely uncorrelated fluctuations.}
        \label{tab:performance_measures_uncorr}
    \end{table}
    
    In the following, we will go into more detail about the results of the demand fluctuations. The results for the solar and wind fluctuations can be found in the appendix \ref{sec:res_fluc_example}.
    
\begin{figure}[H]
        \centering
        \begin{subfigure}[b]{0.49\textwidth}
            \centering
            \includegraphics[width=\textwidth]{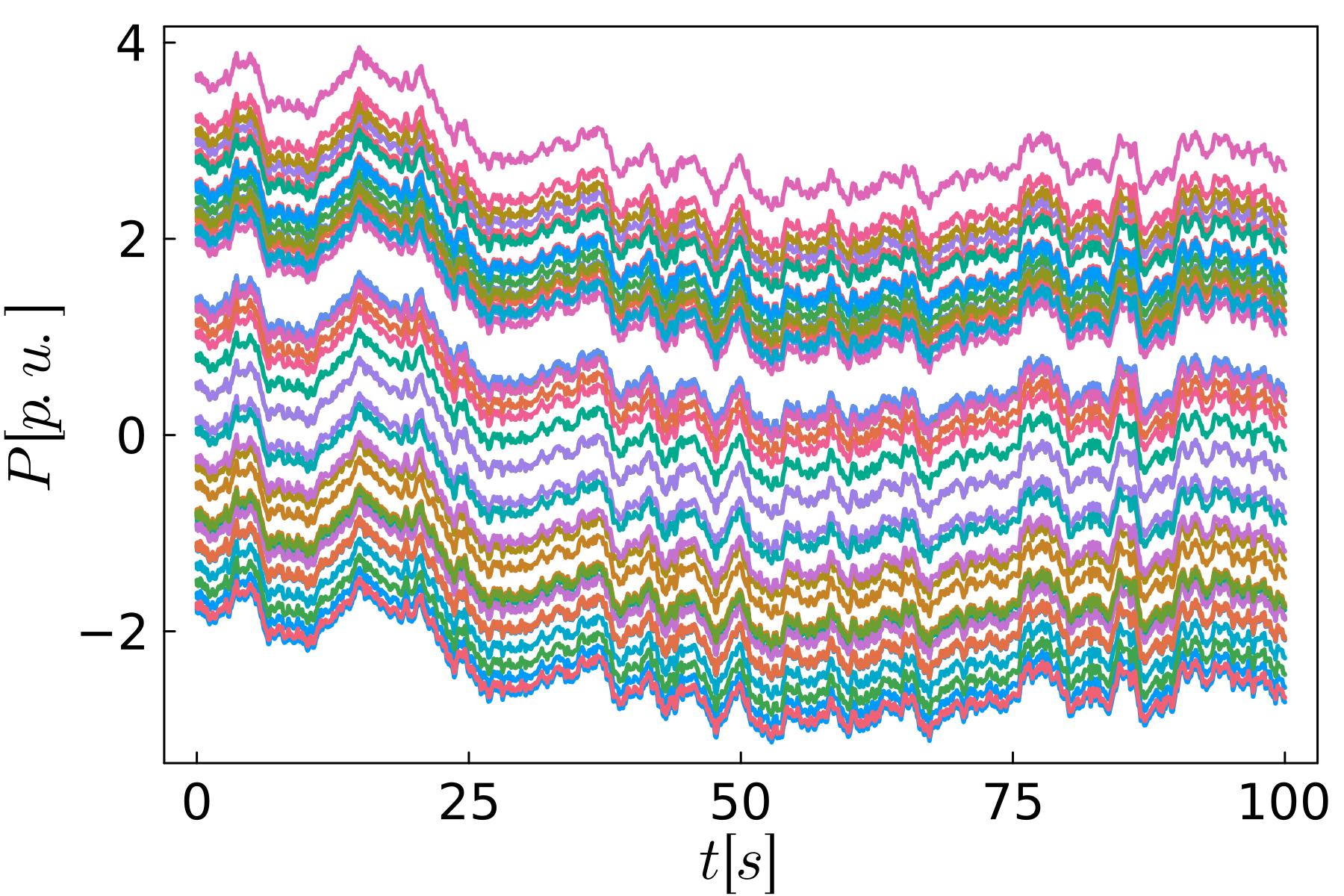}
        \end{subfigure}
        \hfill
        \begin{subfigure}[b]{0.49\textwidth}
            \centering
            \includegraphics[width=\textwidth]{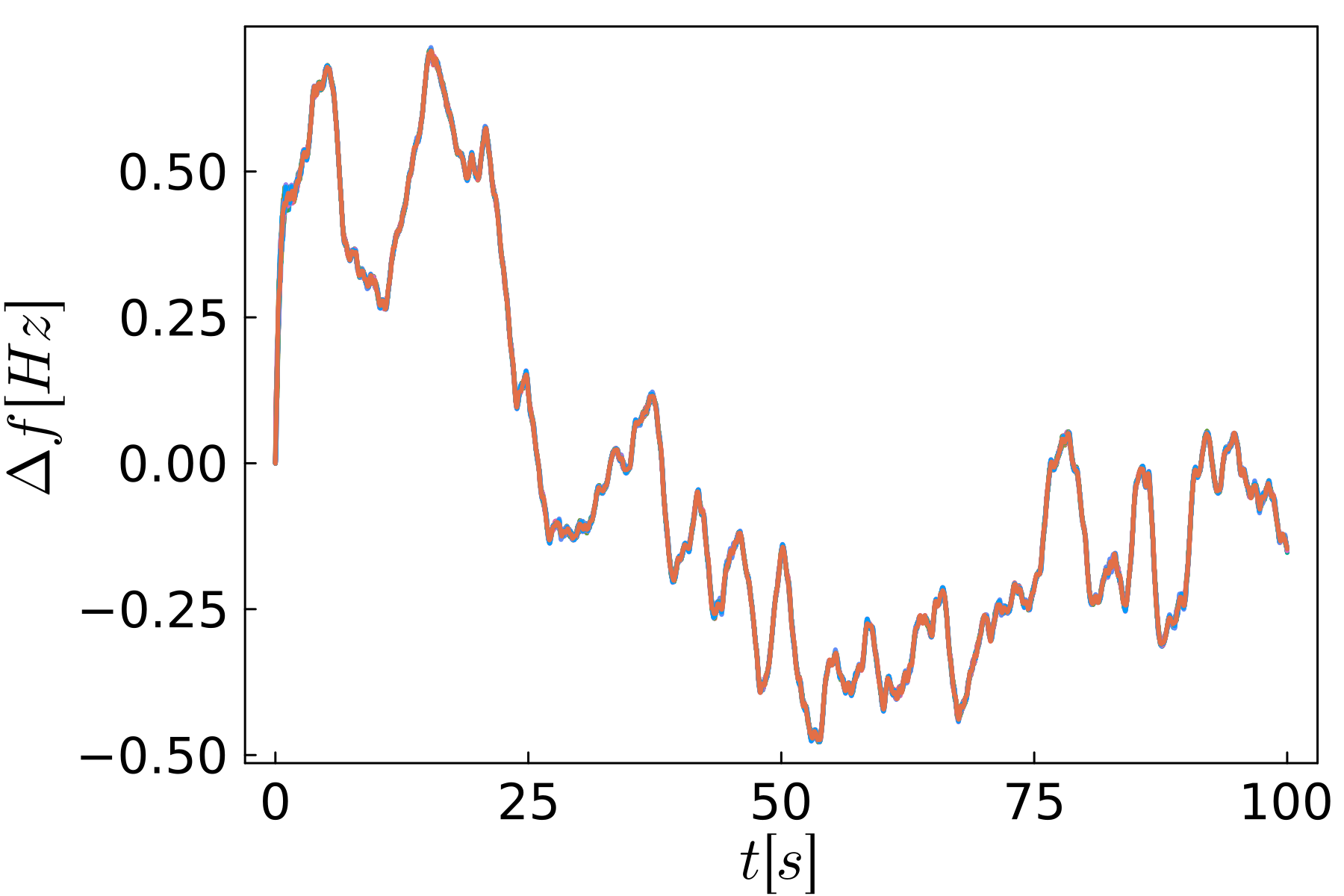}
        \end{subfigure}
        \caption{Results for completely correlated demand fluctuations at the nodes. The figure on the left shows the active powers of the grid-following inverters. The frequency response of the grid-forming inverters is shown in the figure on the right side. The parameters $[\gamma, \epsilon ,\mu_{MB}] = [ 0.016, 33.81, 0.03]$, as in \cite{anvari2022data}, were used to generate the demand fluctuations.}
        \label{fig:demand_fluc_corr}
    \end{figure}

    The figures \ref{fig:demand_fluc_corr} and \ref{fig:demand_fluc_uncorr} show the results for the correlated and uncorrelated demand fluctuations respectively. In this example we use the coefficients for the stochastic process, introduced in \cite{anvari2022data}, which have been extracted from the NOVAREF data set \cite{NOVAREF} which consists of high-resolution demand profiles. In a transmission grid, the number of consumers is significantly higher than in the data sets analyzed in \cite{anvari2022data}. As mentioned in section \ref{sec:demand_fluc} the actual fluctuations should therefore be attenuated. Thus, the result that we present here should be considered a pessimistic estimate. This explains why the frequency response for the uncorrelated fluctuations shown in figure \label{fig:demand_fluc_uncorr} is relatively severe and occasionally even surpasses \SI{0.1}{Hz}. 
    
    For all fluctuation processes considered in this work, we find that the voltage magnitudes of the nodes stay close to the set-point of 1 [p.u.] which is to be expected as we simulate active power fluctuations which couple to the frequency \cite{Power_System_Dynamics_machowski}.

    This example demonstrates that we are able to generate robust and stable synthetic grids. The grid does not lose synchrony even under fluctuations that are stronger than they will be in reality as averaging effects have not been taken into account. 
    
    This opens the door to future research that studies grids that are under severe stress, possibly from compound events, meaning that multiple stressors occur at once. Extreme scenarios that can destabilize grids include the loss of multiple lines, as grids are built N-1 stable, special weather conditions that can cause storage to be locally depleted, causing grid-forming inverters to have to compromise on their grid-forming capabilities and to inject fluctuations as well.
    
    \begin{figure}[H]
        \centering
        \begin{subfigure}[b]{0.49\textwidth}
            \centering
            \includegraphics[width=\textwidth]{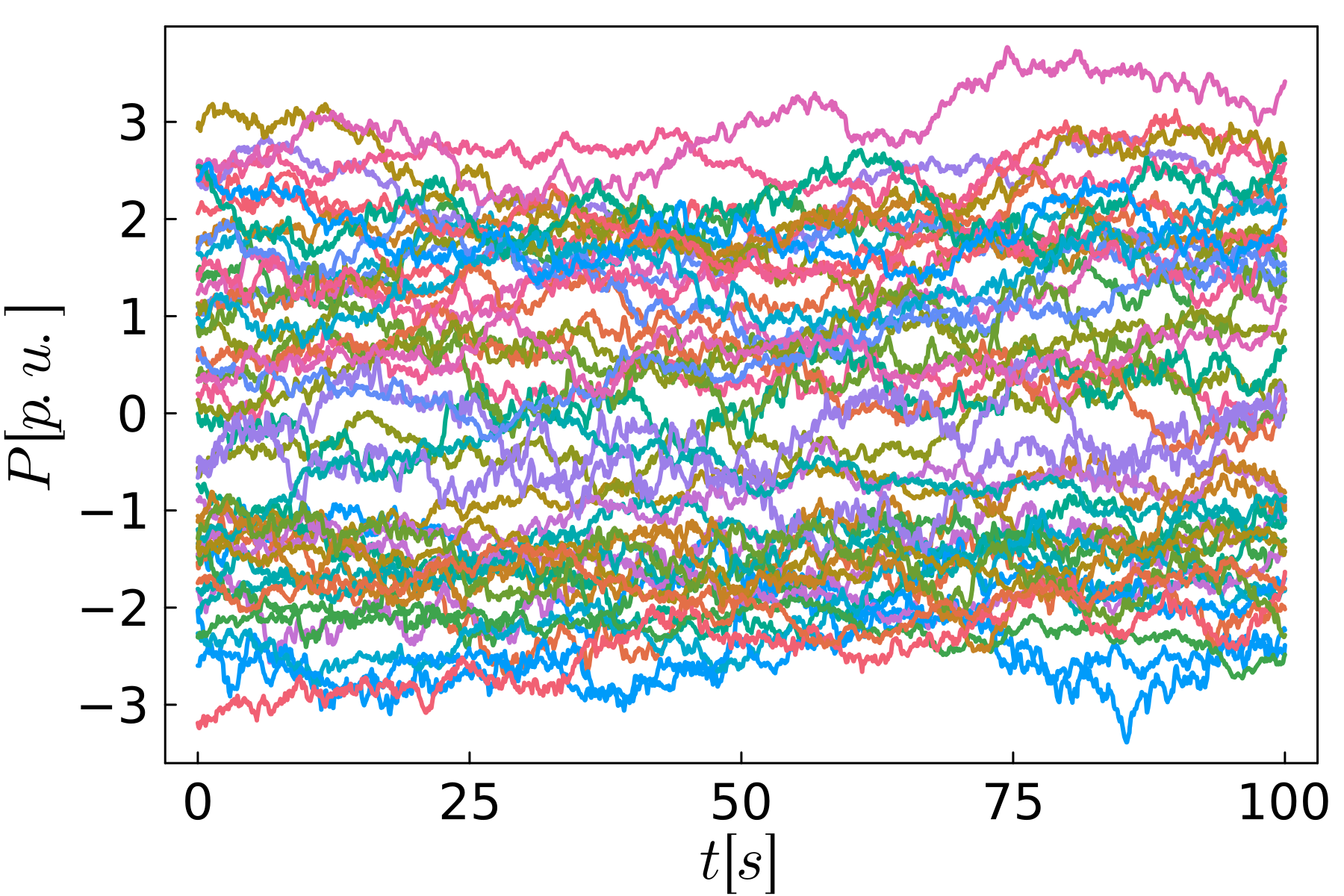}
        \end{subfigure}
        \hfill
        \begin{subfigure}[b]{0.49\textwidth}
            \centering
            \includegraphics[width=\textwidth]{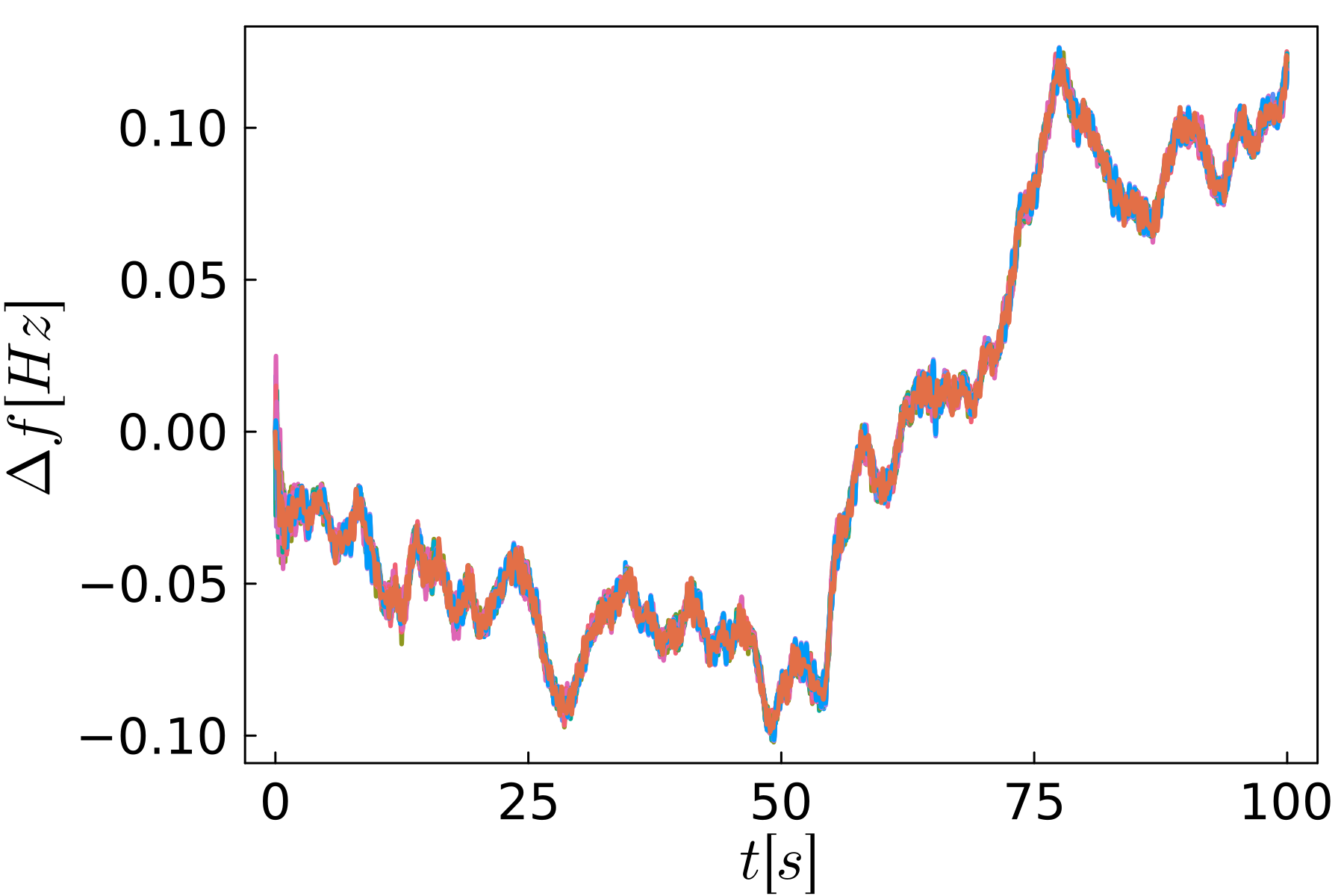}
        \end{subfigure}
        \caption{Results for completely uncorrelated demand fluctuations at the nodes. The figure on the left shows the active powers of the grid-following inverters. The frequency response of the grid-forming inverters is shown in the figure on the right side. The parameters $[\gamma, \epsilon ,\mu_{MB}] = [ 0.016, 33.81, 0.03]$, as in \cite{anvari2022data}, were used to generate the demand fluctuations.}
        \label{fig:demand_fluc_uncorr}
    \end{figure}

 
    \section{Conclusions}
    \label{sec:outlook}
    In this work, a framework to generate synthetic power grid models for studying collective dynamical effects has been introduced. For the first time, the following established methods are combined to obtain synthetic power grids: realistic grid topologies \cite{schultz_random_2014}, active power set-points \cite{Egerer2016Open, taher2019enhancing} and short-term fluctuations, node \cite{kogler_normal_2022} and line models. Finally we introduce validators that ensure our power grid and its operation point fulfill established stability criteria \cite{milano_scripting, transmission_systems_book}, and reject the sample otherwise. Each element in the framework can be substituted as long as it adheres to the general structure thus making the approach modular. For the default elements, we have chosen methods that have already been used and validated in various research projects. We have reviewed these established approaches and draw attention to possible improvements in the respective sections, in particular in order to investigate electricity grids with a high share of renewable energy. We have identified two elements that need improvement, the generation of network topologies and the distribution of active power supply. 
    
    The topologies created with the random growth model \cite{schultz_random_2014} cannot reflect the distribution of transmission line lengths in the empirical \emph{SciGRID} data-set \cite{medjroubi_open_2017}. The model has been designed to resemble network properties, such as the degree distribution, of real EHV power grids. However, the positioning of the nodes is uniformly random, which does not reflect the growth of real power grids. Grid growth is driven by population and demand growth processes that are far from uniform. We assume that it is possible to correct the length distribution by introducing an additional step in the algorithm that considers the geographical location of the nodes. Furthermore, we have assumed that the transmission system topology will remain very similar to today. Future studies should consider how the energy transition influences the topology, as for example, RES are connected to the grid differently than large power plants and the grid evolves to adapt to the new locations.
    
    The major issue in the distribution of active power supply for our synthetic model is that the \emph{ELMOD-DE} \cite{Egerer2016Open} specifies scenarios that reflect the current power supply. As we are interested in studying future dynamics as well, a new method for generating active power distributions is needed. Atlite \cite{Hofmann_2021_atlite} is a software tool that generates weather-dependent power generation potentials and time series for renewable energy technologies. These potentials and time series are promising and could be used to update the active power supply in our model. Further, as the time series depend on the weather, they could also be used to study the synthetic grid under multiple supply scenarios.  
    
    Besides the generation of the synthetic grid dynamics in stable operation points, we also include the major drivers of fluctuations at short time scales. We have implemented the three major drivers of short-term fluctuations in future power grids, solar, wind, and demand. As an example, we study a fully synthetic power grid under these fluctuations. We have decided to add the fluctuations only to the components without grid-forming capabilities as grid-forming components will usually be equipped with sufficient storage. We find that the synthetic grid shows good synchronicity under all three fluctuation scenarios. We saw that there is a relevant contribution to the joint response of synchronous frequencies.
    
    It remains a challenge to find a balance between the simplicity and tractability of the model and realism. We have outlined a wide range of points at which realism can be increased. In the current state, the complete model is already well suited to be used in further research projects. This includes developing methods to study compound and extreme events that particularly stress the system. More immediately it will allow us to advance the study of dynamic power grid stability using graph neural networks ~\cite{nauck2022predicting,nauck_towards_2023, naucktowards2022}. It enables for the first time to generate a large and robust set of heterogeneous DAE models that will challenge the GNN models, and allow us to take one step closer to predicting the dynamic stability of real power grids.
    
    \section*{Data and Code Availability}
    The data that supports the findings of this study are openly available in the GitHub repository \url{https://github.com/PIK-ICoNe/SyntheticPowerGrid_Paper_Companion}. The implementation of the synthetic power grid framework can be found in a separate GitHub repository \url{https://github.com/PIK-ICoNe/SyntheticPowerGrids.jl}.
    
    \section*{Acknowledgments}
    A. Büttner acknowledges support from the German Academic Scholarship Foundation. 
    This research is partially funded within the framework of the "Technology-oriented systems analysis" funding area of the BMWi’s 7th Energy Research Program "Innovation for the energy transition" (FKZ: 03EI1016B) and the DFG-project (ExSyCo-Grid, 410409736), as well as the DFG-project (CoComusy, Grant No. KU 837/39-1 / RA 516/13-1).

    \section*{References}
    
    \bibliographystyle{iopart-num}
    \bibliography{main}

\section*{Appendix}

    \subsection*{Line Length Distribution}
    \begin{figure}[H]
        \centering
        \includegraphics[width=0.7\textwidth]{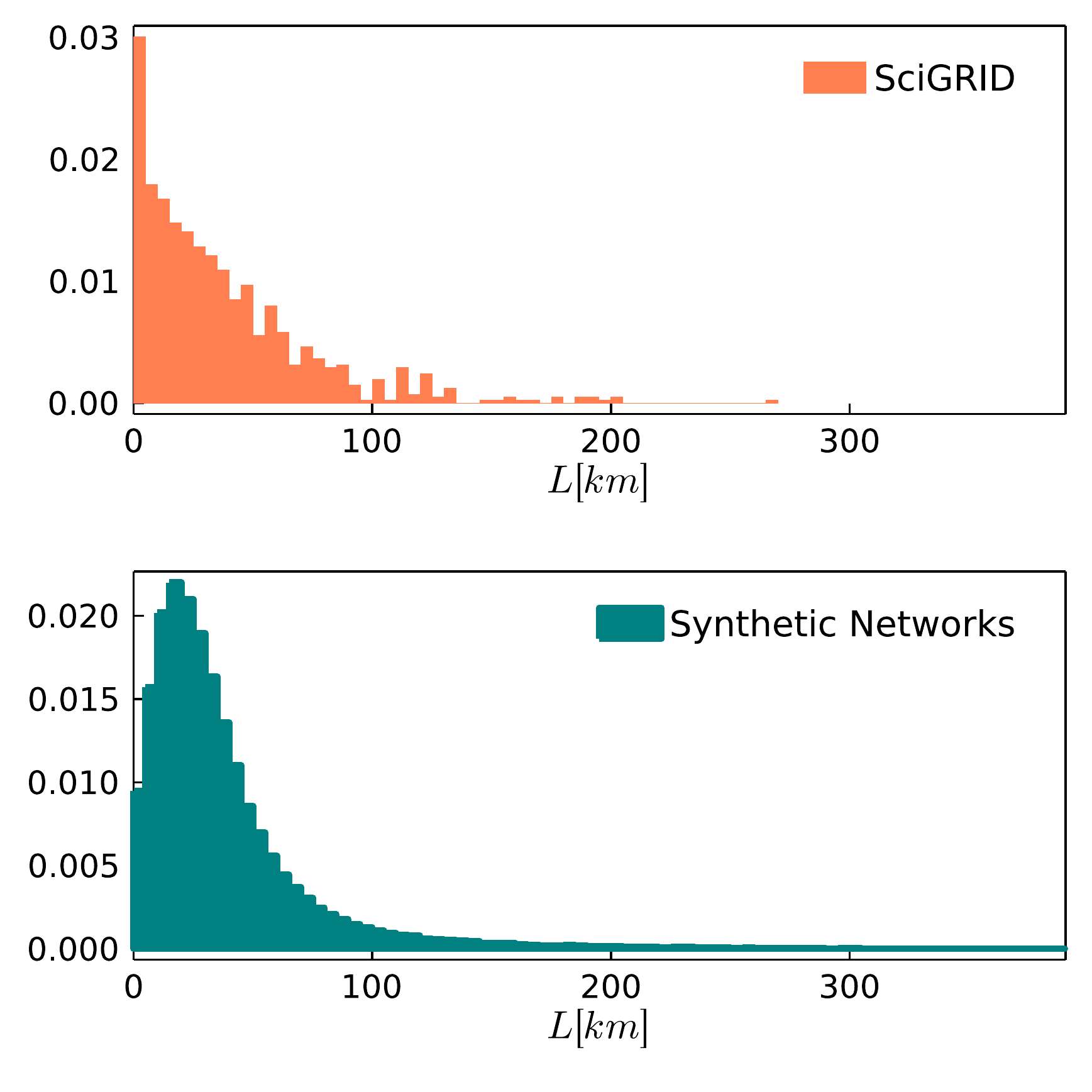}
        \caption{Histograms of the line lengths in the \emph{SciGRID} data set \cite{medjroubi_open_2017} and of our synthetic model. Both distributions show heavy tails. The data for the \emph{SciGRID} lines indicates a scale-free distribution but the quantity of data is too small to make accurate statements. Further investigations are necessary.}
        \label{fig:line_lengths}
    \end{figure}
    
    \subsection*{RES Fluctuation Examples}
    \label{sec:res_fluc_example}
    \begin{figure}[H]
        \centering
        \begin{subfigure}[b]{0.49\textwidth}
            \centering
            \includegraphics[width=\textwidth]{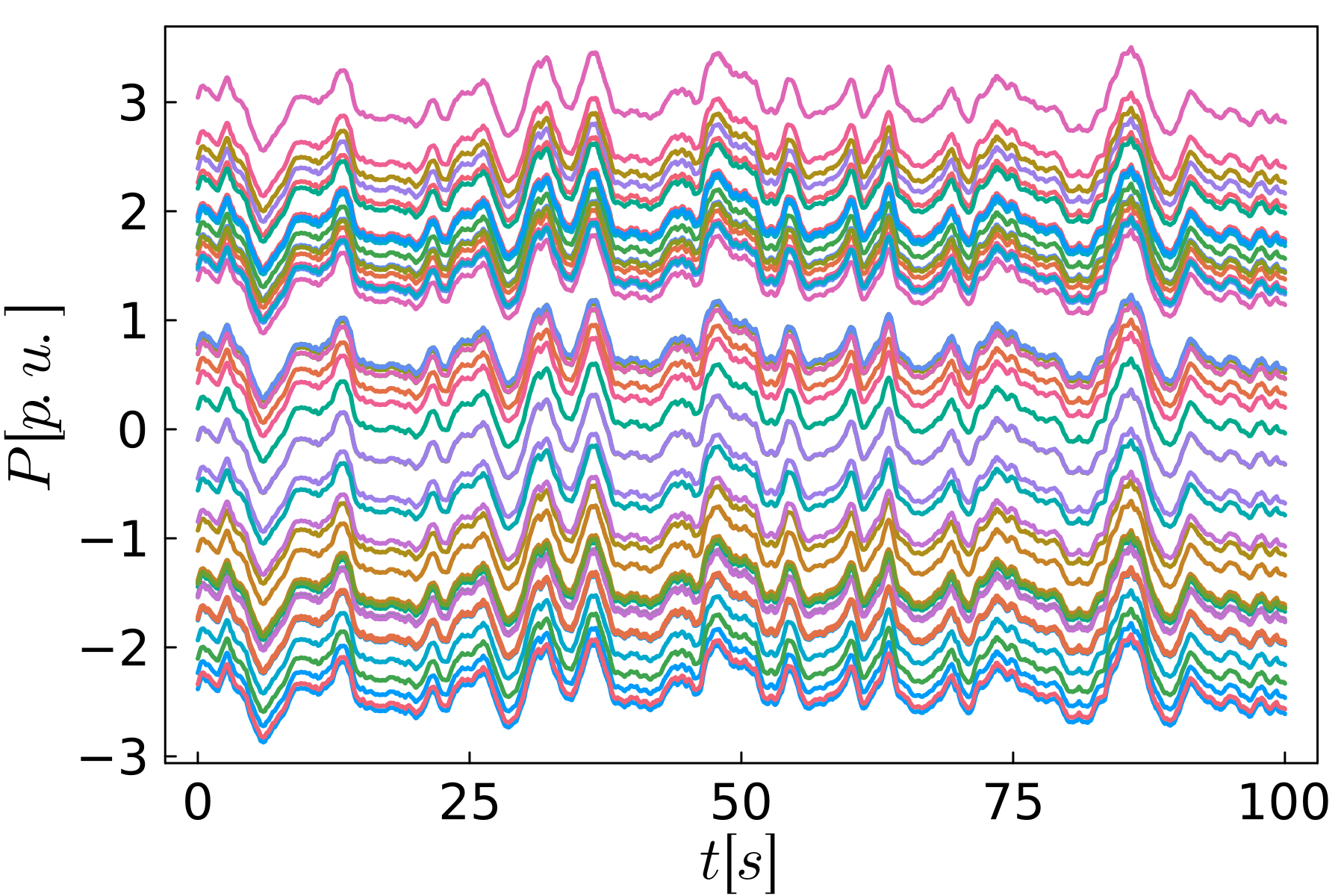}
        \end{subfigure}
        \hfill
        \begin{subfigure}[b]{0.49\textwidth}
            \centering
            \includegraphics[width=\textwidth]{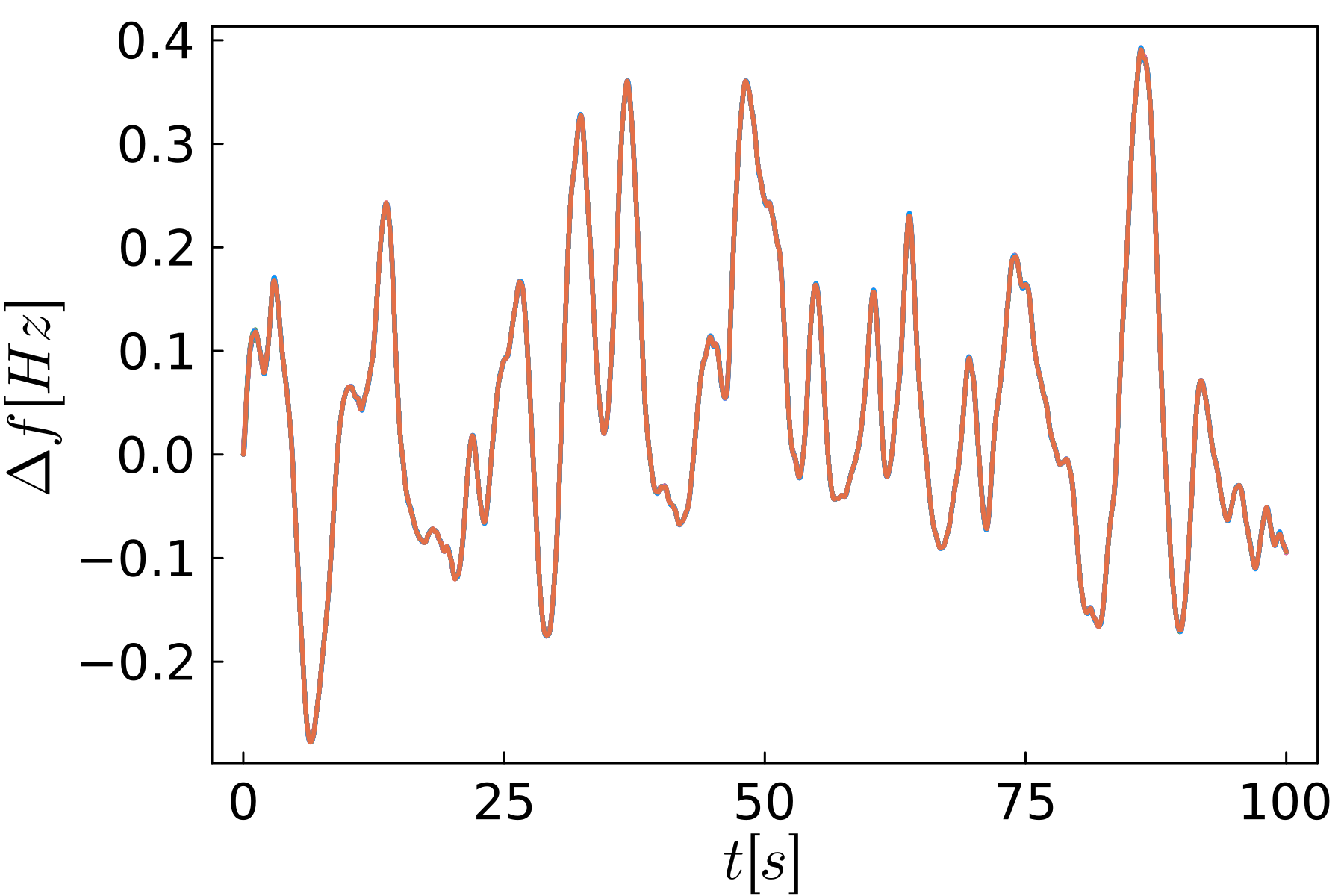}
        \end{subfigure}
        \caption{Results for completely correlated wind power fluctuations. The parameters $[D, \gamma, g, \epsilon] = [0.1, 1.0, 0.5, 1.0 ]$, as in \cite{schmietendorf2017impact}, were used to generate the wind power fluctuations.}
    \end{figure}
    
    \begin{figure}[H]
        \centering
        \begin{subfigure}[b]{0.49\textwidth}
            \centering
            \includegraphics[width=\textwidth]{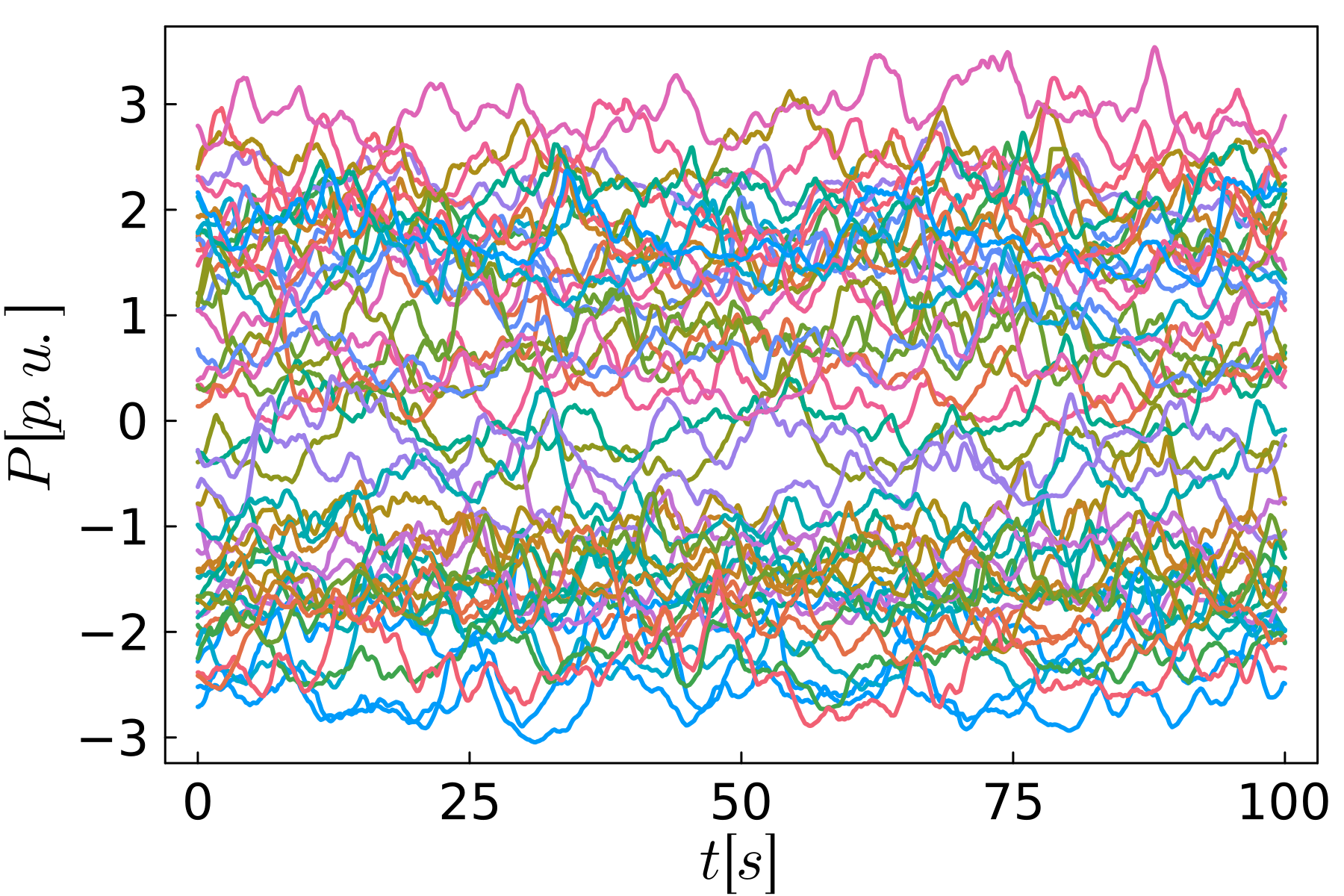}
         \end{subfigure}
         \hfill
        \begin{subfigure}[b]{0.49\textwidth}
            \centering
            \includegraphics[width=\textwidth]{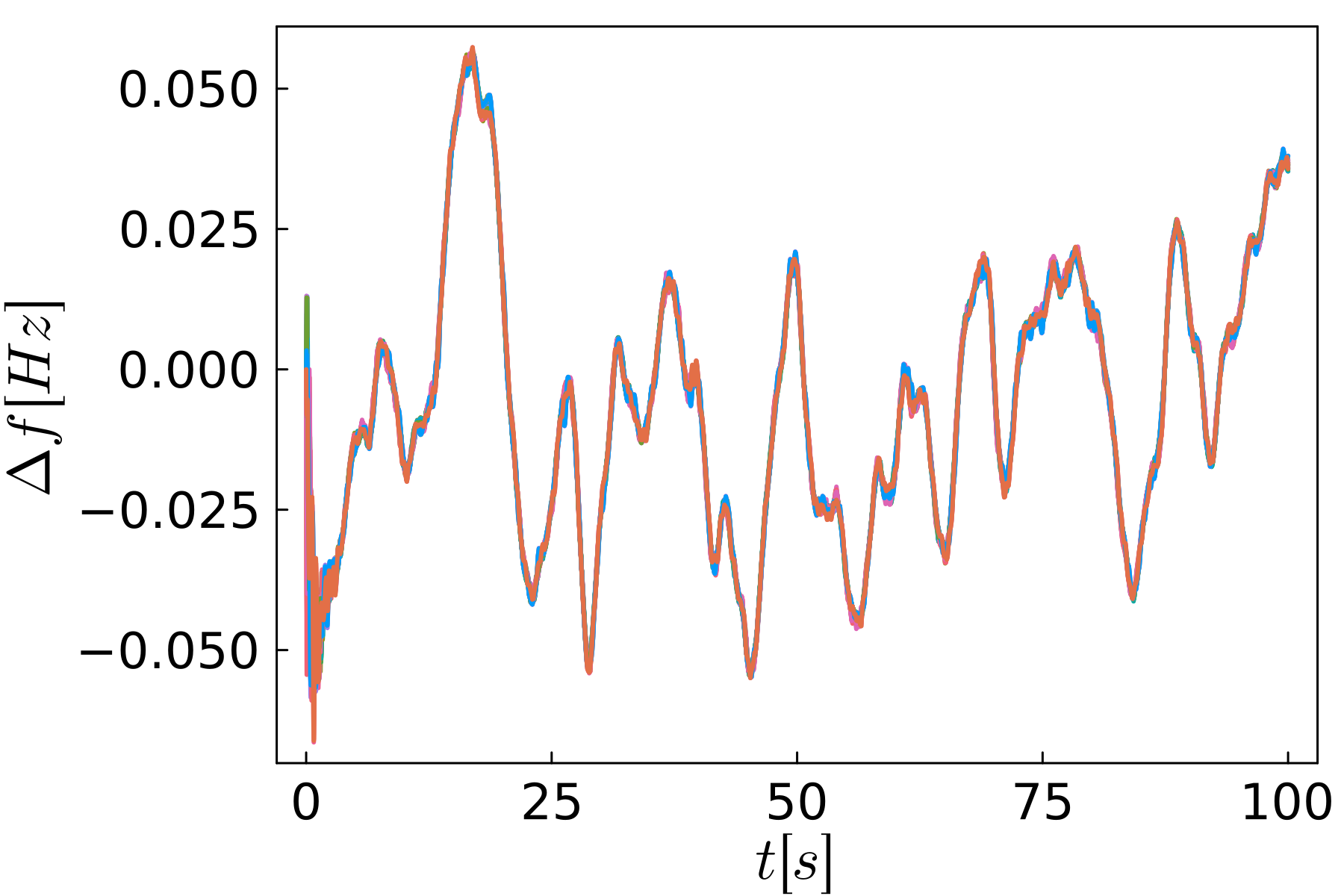}
        \end{subfigure}
        \caption{Results for completely uncorrelated wind power fluctuations. The parameters $[D, \gamma, g, \epsilon] = [0.1, 1.0, 0.5, 1.0 ]$, as in \cite{schmietendorf2017impact}, were used to generate the wind power fluctuations.}
    \end{figure}
    
    \begin{figure}[H]
        \centering
        \begin{subfigure}[b]{0.49\textwidth}
            \centering
            \includegraphics[width=\textwidth]{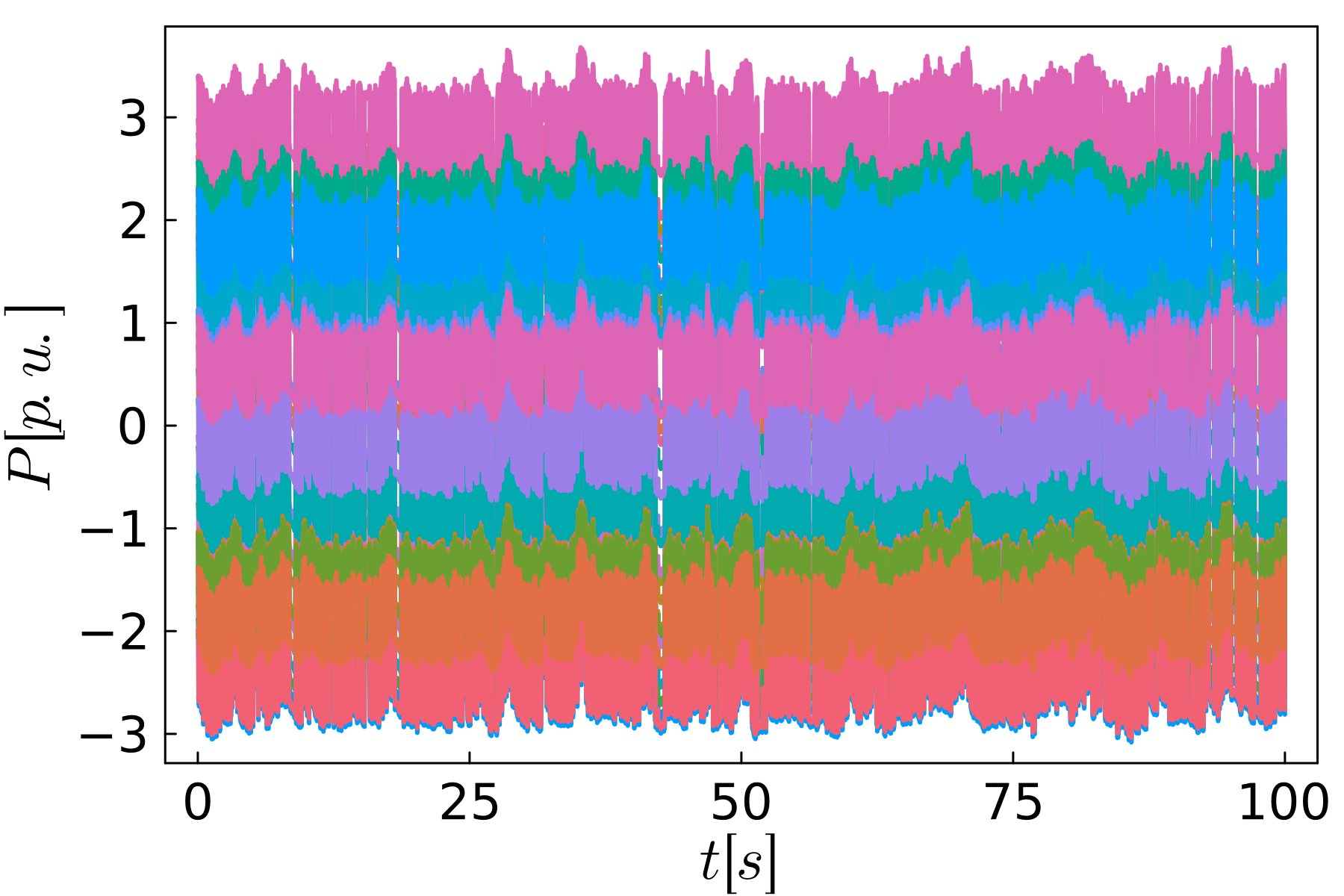}
        \end{subfigure}
        \hfill
        \begin{subfigure}[b]{0.49\textwidth}
            \centering
            \includegraphics[width=\textwidth]{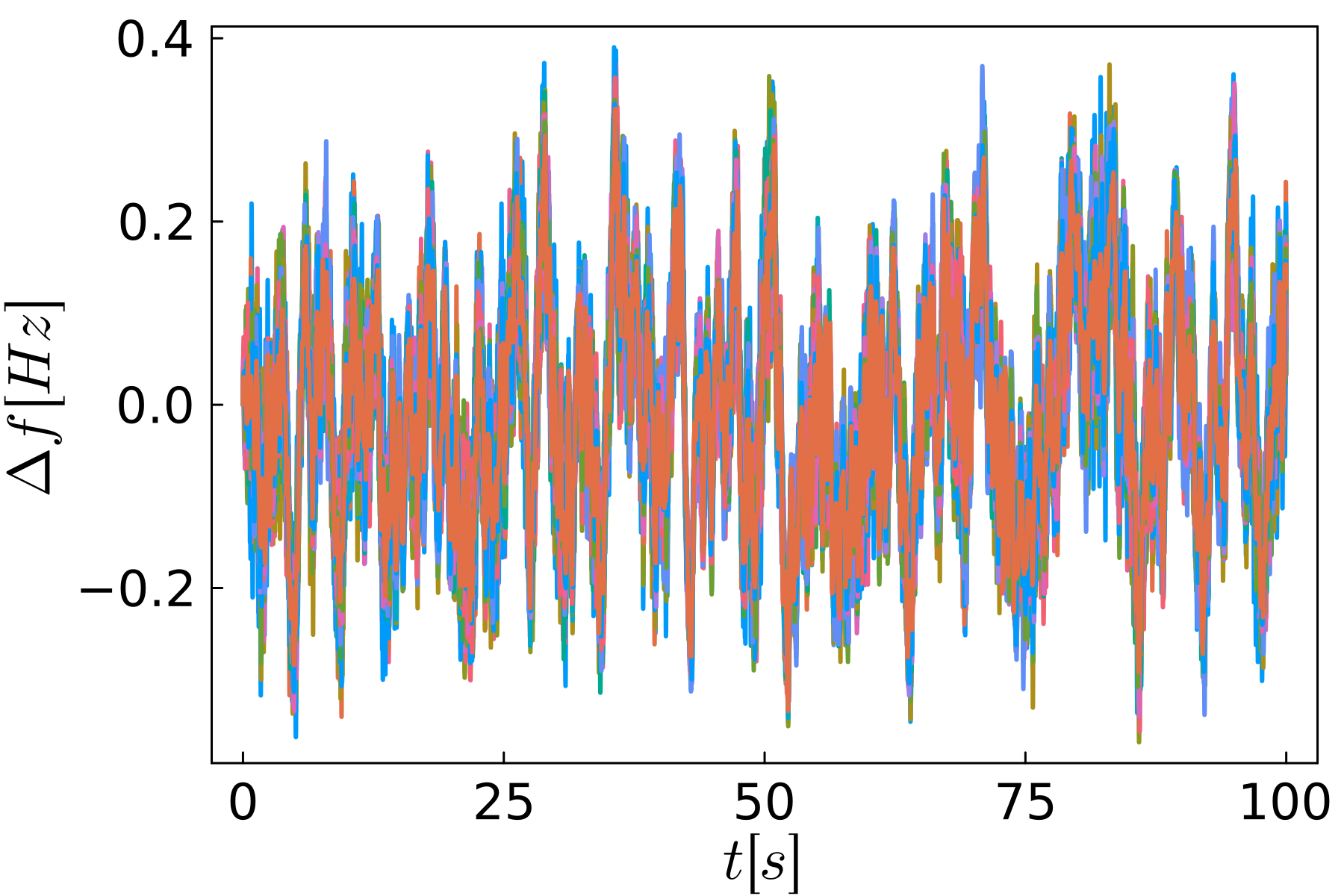}
        \end{subfigure}
        \caption{Results for completely correlated solar power fluctuations.  The parameters $[D^{(2)}, \lambda, \sigma_{\eta}] = [0.001, 0.01, 0.02]$, as in \cite{anvari2017suppressing}, were used to generate the solar power fluctuations.}
    \end{figure}
    
    \begin{figure}[H]
        \centering
        \begin{subfigure}[b]{0.49\textwidth}
            \centering
            \includegraphics[width=\textwidth]{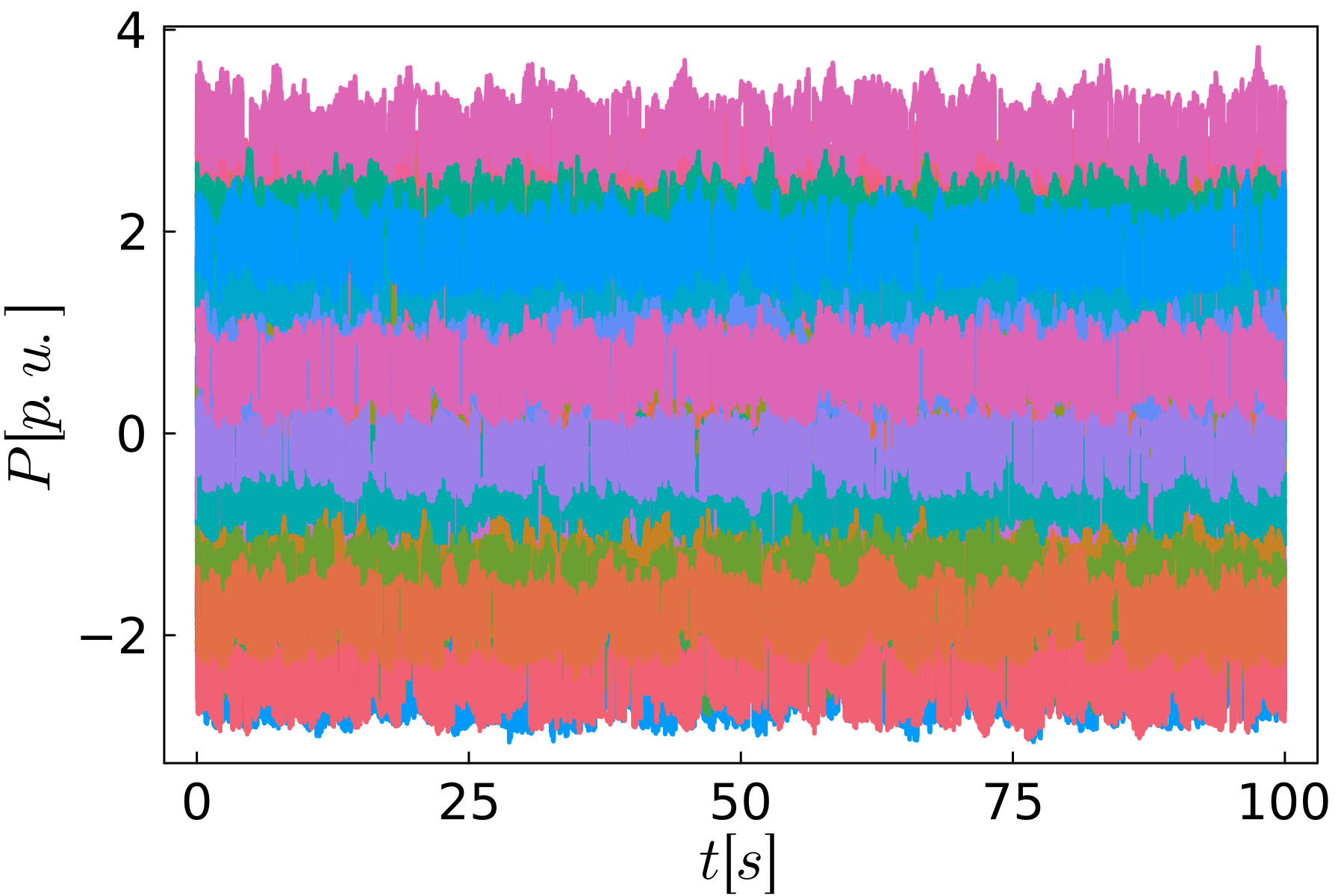}
         \end{subfigure}
         \hfill
        \begin{subfigure}[b]{0.49\textwidth}
            \centering
            \includegraphics[width=\textwidth]{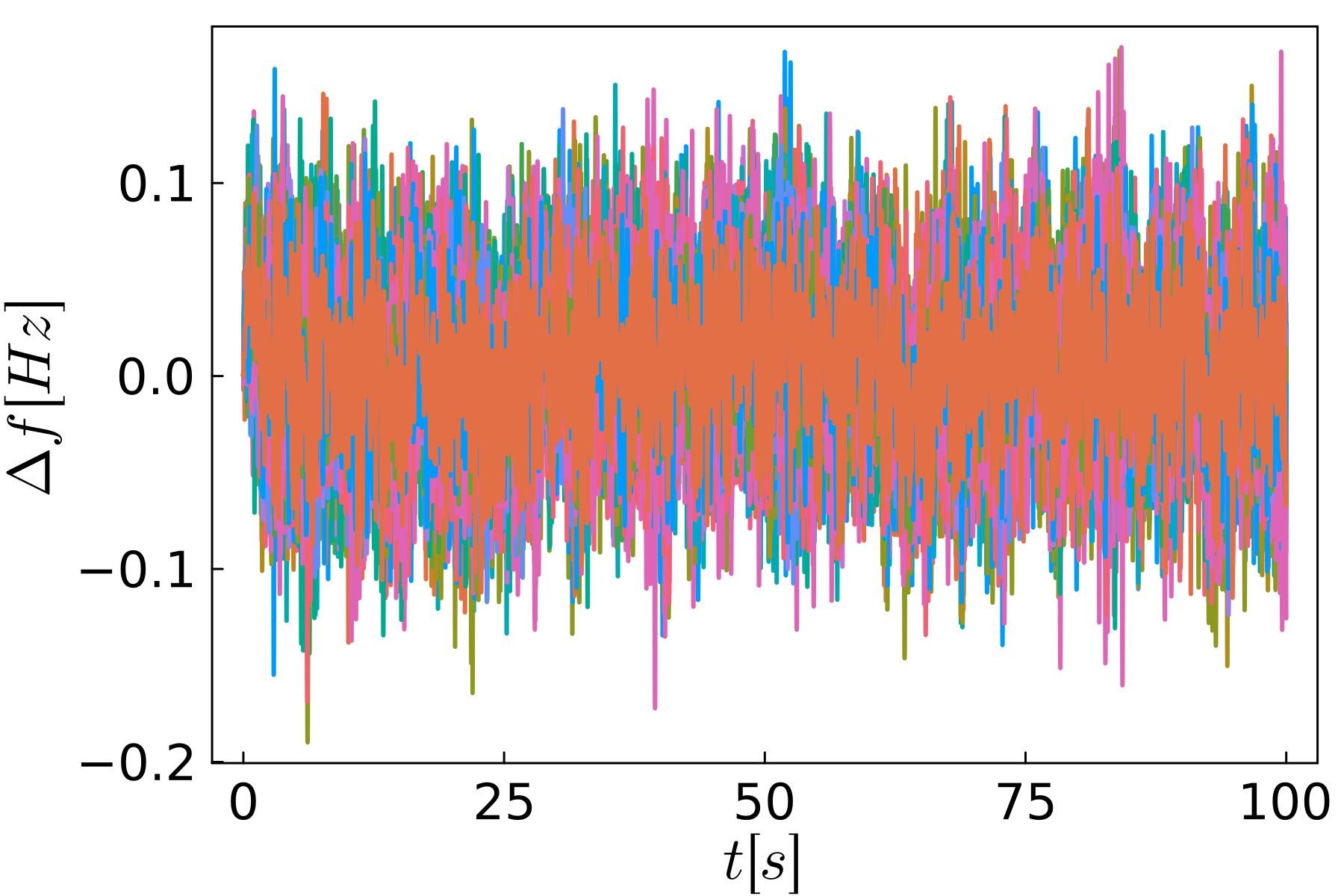}
        \end{subfigure}
        \caption{Results for completely uncorrelated solar power fluctuations.  The parameters $[D^{(2)}, \lambda, \sigma_{\eta}] = [0.001, 0.01, 0.02]$, as in \cite{anvari2017suppressing}, were used to generate the solar power fluctuations.}
    \end{figure}
\end{document}